\documentclass{pasa}%

\usepackage{graphicx}
\usepackage{epsfig}
\usepackage{epstopdf}
\usepackage{subfigure}
\usepackage{amssymb}
\usepackage{rotating}
\usepackage{textcomp}
\usepackage[usenames,dvipsnames]{color}
\usepackage{float,lscape}
\usepackage[flushleft]{threeparttable} 
\newcommand{\farcm}{\mbox{\ensuremath{.\mkern-4mu^\prime}}}

\newcommand{\arcdeg}{\ensuremath{^{\circ}}}
\makeatletter

\title[Cross-identifying the GLEAM 4-Jy Sample]{The GLEAM 4-Jy (G4Jy) Sample: \newline II. Host-galaxy identification for individual sources}

\author[Sarah V. White et al.]{Sarah V. White$^{1, 2}$\thanks{sarahwhite.astro@gmail.com}, Thomas M.O. Franzen$^{1, 3}$, Chris J. Riseley$^{4,5,6}$, O. Ivy Wong$^{7}$, \newline Anna D. Kapi{\'n}ska$^{7,8}$, Natasha Hurley--Walker$^{1}$, Joseph R. Callingham$^{3}$, Kshitij Thorat$^{2,9}$, \newline Chen Wu$^{7}$, Paul Hancock$^{1}$, Richard W. Hunstead$^{10}$, Nick Seymour$^{1}$, Jesse Swan$^{11}$, Randall Wayth$^{1}$, John Morgan$^{1}$, Rajan Chhetri$^{1}$, Carole Jackson$^{3}$, Stuart Weston$^{12}$, Martin Bell$^{13}$, B.\,M. Gaensler$^{14}$, Melanie Johnston--Hollitt$^{1}$, Andr{\'e} Offringa$^{3}$, and Lister Staveley--Smith$^{7}$ 
\affil{$^{1}$International Centre for Radio Astronomy Research (ICRAR), Curtin University,  Bentley, WA 6102, Australia}%
\affil{$^{2}$Department of Physics and Electronics, Rhodes University, PO Box 94, Grahamstown, 6140, South Africa}%
\affil{$^{3}$ASTRON: the Netherlands Institute for Radio Astronomy, Oude Hoogeveensedijk 4, 7991 PD, Dwingeloo, The Netherlands}%
\affil{$^{4}$CSIRO Astronomy and Space Science, PO Box 1130, Bentley, WA 6102, Australia}%
\affil{$^{5}$Dipartimento di Fisica e Astronomia, Universit\`a degli Studi di Bologna, via P. Gobetti 93/2, 40129 Bologna, Italy}%
\affil{$^{6}$INAF -- Istituto di Radioastronomia, via P. Gobetti 101, 40129 Bologna, Italy}%
\affil{$^{7}$ICRAR, University of Western Australia (M468), 35 Stirling Highway, Crawley, WA 6009, Australia}%
\affil{$^{8}$National Radio Astronomy Observatory (NRAO), 1003 Lopezville Rd, Socorro NM 87801, USA}%
\affil{$^{9}$South African Radio Astronomy Observatory (SARAO), 2 Fir Street, Observatory, Cape Town, 7925, South Africa}%
\affil{$^{10}$Sydney Institute for Astronomy (SIfA), School of Physics, University of Sydney, NSW 2006, Australia}%
\affil{$^{11}$School of Physical Sciences, University of Tasmania, Private Bag 37, Hobart, Tasmania, 7001 Australia}%
\affil{$^{12}$Institute for Radio Astronomy and Space Research (IRASR), Auckland University of Technology, Auckland 1010, New Zealand}%
\affil{$^{13}$University of Technology Sydney, 15 Broadway, Ultimo NSW 2007, Australia}%
\affil{$^{14}$Dunlap Institute for Astronomy and Astrophysics, University of Toronto, Toronto, ON M5S 3H4, Canada}%

}%

\jid{PASA}
\doi{10.1017/pasa.\the\year.xxx}
\jyear{\the\year}

\usepackage{aas_macros}
\usepackage{hyperref} 
\hypersetup{colorlinks,citecolor=blue,linkcolor=blue,urlcolor=blue}

\hypersetup{draft}

\begin{document}

\begin{frontmatter}
\maketitle

\begin{abstract}
The entire southern sky (Declination, $\delta< 30^{\circ}$) has been observed using the Murchison Widefield Array (MWA), which provides radio imaging of $\sim$2$'$ resolution at low frequencies (72--231\,MHz). This is the GaLactic and Extragalactic All-sky MWA (GLEAM) Survey, and we have previously used a combination of visual inspection, cross-checks against the literature, and internal matching to identify the `brightest' radio-sources ($S_{\mathrm{151\,MHz}}>4$\,Jy) in the extragalactic catalogue (Galactic latitude, $|b| >10^{\circ}$). We refer to these 1,863 sources as the GLEAM 4-Jy (G4Jy) Sample, and use radio images (of $\leq45''$ resolution), and multi-wavelength information, to assess their morphology and identify the galaxy that is hosting the radio emission (where appropriate). Details of how to access all of the overlays used for this work are available at https://github.com/svw26/G4Jy. Alongside this we conduct further checks against the literature, which we document in this paper for individual sources. Whilst the vast majority of the G4Jy Sample are active galactic nuclei with powerful radio-jets, we highlight that it also contains a nebula, two nearby, star-forming galaxies, a cluster relic, and a cluster halo. There are also three extended sources for which we are unable to infer the mechanism that gives rise to the low-frequency emission. In the G4Jy catalogue we provide mid-infrared identifications for 86\% of the sources, and flag the remainder as: having an uncertain identification (129 sources), having a faint/uncharacterised mid-infrared host (126 sources), or it being inappropriate to specify a host (2 sources). For the subset of 129 sources, there is ambiguity concerning candidate host-galaxies, and this includes four sources (B0424$-$728, B0703$-$451, 3C~198, and 3C~403.1) where we question the existing identification. 
\end{abstract}

\begin{keywords}
catalogues -- galaxies: active -- galaxies: evolution -- radio continuum: galaxies \newline

(Received 15 October 2019; revised 23 March 2020; accepted 23 March 2020)
\end{keywords}
\end{frontmatter}

\section{INTRODUCTION}
\label{sec:intro}

Active galactic nuclei (AGN) are believed to influence the way in which their host galaxies evolve, through thermal feedback \citep[e.g.][]{Croton2006} connected with the accretion process, and kinetic feedback \citep[e.g.][]{Gaibler2012} associated with powerful, relativistic radio-jets. However, the overall impact of the latter is debated. In some cases radio jets may {\it suppress} star formation by expelling gas from the system \citep[e.g.][]{Morganti2013}, whilst in others they appear to {\it promote} star formation by triggering the collapse of a molecular cloud \citep[e.g.][]{Croft2006}. In order to appreciate which scenario is more significant, a large sample of `jetted' AGN is required in order to determine robust statistics on their properties, and to account for factors such as the size of the molecular-gas reservoir \citep{Emonts2011}, the jet power \citep{Mukherjee2016}, and the inclination of the jet with respect to the galaxy plane \citep{GarciaBurillo2014}.

In addition, a large sample is needed for disentangling the effects of jet power, age and environmental density on the radio luminosity, and exploring these properties as a function of redshift \citep{Wang2008,Hardcastle2018}. Furthermore, we are yet to understand the actual mechanism by which material, accreting onto the central supermassive black-hole, is launched as collimated jets. It is thought that magnetic fields play a crucial role in this process \citep{Blandford1977,Blandford1982,Tchekhovskoy2011}, and so polarimetric observations offer an excellent opportunity for further investigation \citep[e.g.][]{Hovatta2019}. For this, and high-resolution follow-up using very long baseline interferometry (VLBI), it is necessary to use a sample of bright radio-sources detected at a high signal-to-noise ratio.   

We have defined such a sample using observations from the Murchison Widefield Array \citep[MWA;][]{Tingay2013} over the entire southern sky (Declination, $\delta< 30^{\circ}$). This is the GaLactic and Extragalactic All-sky MWA \citep[GLEAM;][]{Wayth2015} Survey, and we use the extragalactic catalogue \citep[Galactic latitude, $|b| >10^{\circ}$;][]{HurleyWalker2017} to identify 1,863 sources with an integrated flux-density greater than 4\,Jy at 151\,MHz. These we collectively refer to as the GLEAM 4-Jy (G4Jy) Sample \citep{Jackson2015,White2018}, for which full-sample details are described in the accompanying definition paper (\citealt{White2020a}; hereafter `Paper~I'). This sample is over 10 times larger than that of the revised Third Cambridge Catalogue of Radio Sources \citep[3CRR;][]{Laing1983}, which is also selected at low frequencies ($S_{\mathrm{178\,MHz}}>10.9$\,Jy). As a result, both samples avoid the orientation bias that is inherent in higher-frequency radio surveys, where Doppler boosting \citep{Rees1966,Blandford1979} leads to a greater-than-expected fraction of radio AGN with their jet axis close to the line-of-sight. 

Having constructed a sample of bright, extragalactic radio-sources, based on new `blind' survey data, the next important step is to determine the host galaxy of the radio emission for each source. This is necessary if the low-frequency radio information is to be combined with other datasets, allowing us to build a comprehensive, multi-wavelength view of the different processes taking place within these objects. However, cross-identification is notoriously difficult for extended radio sources and those with complex morphology, as demonstrated by the Radio Galaxy Zoo project \citep{Banfield2015}\footnote{\citet{Banfield2015} ``find that the source identifications from Radio Galaxy Zoo volunteers are as likely to disagree as the experts for difficult or `unusual' radio sources''.}. It is also further complicated by: (i) the resolution of the radio images \citep{McAlpine2012,Tang2019}, (ii) projection effects \citep{Reynolds1980}, (iii) the depth of the radio data \citep{Smolcic2017}, (iv) the depth of optical/infrared data used for the identification \citep{McAlpine2012}, and (v) there being different mechanisms that can produce the same radio morphology \citep[e.g.][]{Jones1992}. Whilst effort is being invested in developing machine-learning algorithms for morphology classification \citep[e.g.][]{Aniyan2017,Wu2019} and likelihood-ratio methods for cross-identification \citep[e.g.][]{Weston2018,Williams2019}, visual inspection still remains the most-reliable method. Indeed, this is the method used by \citet[][]{Williams2019} for classifying and identifying the subset of their sources that are large, bright, and otherwise complex in morphology. Hence, visual inspection is the approach that we use for cross-identifying the G4Jy Sample, with 1,863 being a feasible number of sources to consider.

\subsection{Paper outline}

The previous G4Jy paper, Paper~I, explains how the G4Jy Sample is constructed, and presents the value-added catalogue for these bright radio-sources. In this paper, Paper~II, we provide an outline of the multi-wavelength data used for defining and inspecting the sample (Section~\ref{sec:observations}), as well as a guide to the labels/flags recorded in the catalogue that are most-relevant for host-galaxy identification (Section~\ref{sec:relevantflags}). Further details can be found in Paper~I, and we emphasise that all overlays/images for the full sample are available online\footnote{Please see https://github.com/svw26/G4Jy for details of how to download them.}. Section~\ref{sec:variety} shows the wide variety of radio sources in the sample (not all of which are AGN), and additional literature checks are documented in Section~\ref{sec:additionalchecks}. A summary of the findings presented in this paper is in Section~\ref{sec:finalsummary}.

Unless otherwise specified, we use integrated flux-densities (as opposed to peak surface-brightnesses). In addition, we use a $\Lambda$CDM cosmology, with $H_{0} = 70$\,km\,s$^{-1}$\,Mpc$^{-1}$, $\Omega_{m}=0.3$, $\Omega_{\Lambda}=0.7$. Source names that are based on B1950 co-ordinates are indicated via the prefix `B', whilst all other position-derived names refer to J2000 co-ordinates. The sign convention that we use for a spectral index, $\alpha$, is as defined by $S_{\nu} \propto \nu^{\alpha}$.

\section{An overview of the data} \label{sec:observations}

Identifying the host galaxy of a radio source goes hand-in-hand with understanding its radio morphology. Although the GLEAM Survey (Dec.\,$<30^{\circ}$) provides excellent spectral coverage and sensitivity to diffuse emission at low frequencies, it has poor spatial resolution ($\sim$2$'$). Therefore, we need to consider information from other radio surveys (with better spatial resolution) in order to pinpoint where the radio emission originates. In many cases the positional distribution of candidate host-galaxies, with respect to the radio emission, is also informative. For assessing these candidates, optical images have historically been used \citep[e.g.][]{Jones1992}, but these introduce a bias against galaxies that are dust-obscured. Therefore, we choose to identify host galaxies in the mid-infrared instead, since warm dust (heated by either star formation or an AGN) radiates at these wavelengths. 

For our visual inspection of the brightest radio sources in GLEAM, we overlay multiple sets of radio contours onto mid-infrared images. We create one set of overlays that are 1$^{\circ}$ across, for considering large-scale, extended radio-emission, and another set of overlays that are 10$'$ across, for identifying the host galaxy in the mid-infrared. For both sets we use each of the following datasets:

\begin{itemize}
\item{The extragalactic catalogue (EGC) of the GLEAM Survey \citep{HurleyWalker2017}, which includes measurements over 72--231\,MHz. For visual inspection, we use the wide-band (170--231\,MHz) images, which have a spatial resolution of $\sim$2$'$.}
\item{The TIFR GMRT Sky Survey (TGSS) first alternative data release (ADR1) catalogue and images \citep{Intema2017}, at 150\,MHz. This survey provides 25$''$ resolution for sources at Dec.\,$>-53^{\circ}$. }
\item{The Sydney University Molonglo Sky Survey (SUMSS) catalogue and images \citep{Mauch2003,Murphy2007}, at 843\,MHz. These radio images have a spatial resolution of $45'' \times 45''\ \mathrm{cosec} |\delta|$, at position angle = 0$^{\circ}$.}
\item{The NRAO (National Radio Astronomy Observatory) VLA (Very Large Array) Sky Survey \citep[NVSS;][]{Condon1998} catalogue and images, at 1.4\,GHz. The resolution of these images is 45$''$.}
\item{Positions from the Australia Telescope 20-GHz (AT20G) catalogue \citep{Murphy2010}.}
\item{The All-sky {\it Wide-field Infrared Survey Explorer} (AllWISE) catalogue \citep{Cutri2013}, with mid-infrared data at 3.4, 4.6, 12, and 22\,$\mu$m. We use the 3.4-$\mu$m (`W1' band) images for visual inspection, which have a resolution of 6.1$''$.}
\item{Positions from the 6-degree Field Galaxy Survey (6dFGS) optical catalogue \citep{Jones2004}.}
\end{itemize}

Note that NVSS is used for sources at Dec. $\geq-39.5^{\circ}$, whilst SUMSS is used for sources below this declination. For further details regarding each of the above surveys and how we use them, see sections~2, 4, and 5 of Paper~I.

\section{Morphology labels and flags related to visual inspection} \label{sec:relevantflags}

Full details regarding our visual inspection are presented in section~5 of Paper~I, so here (for quick reference) we provide a reminder of labels and flags that we assign to the G4Jy sources during this process. Specifically, those concerning radio morphology, potential source confusion, and the host galaxy itself, as these are the most pertinent for host-galaxy identification.

\subsection{Morphology classification}
\label{sec:morphology}

When inspecting each GLEAM component considered for the G4Jy Sample, we use the radio contours in TGSS and/or NVSS/SUMSS for classifying the morphology. The four categories that we use are:

\begin{itemize}
\item `single' -- the source has a simple (typically compact) morphology in TGSS and NVSS/SUMSS;
\item `double' -- the source has two lobes evident in TGSS or NVSS/SUMSS, but there is no distinct detection of a core; {\it or} it has an elongated structure that is suggestive of lobes, but is accompanied by a single, catalogued detection;
\item `triple' -- the source has two lobes evident in TGSS or NVSS/SUMSS, and there is a distinct detection of a core in the same survey;
\item `complex' -- the source has a complicated morphology that does not clearly belong to any of the above categories.
\end{itemize} 

We find that the G4Jy Sample contains 1,245 `single' sources, 479 `doubles', 77 `triples', and 62 sources with `complex' morphology.

\subsection{Flag to indicate source confusion}

The low-frequency flux-densities for a G4Jy source will be overestimates if there are other, unrelated sources that are blended together by the MWA beam. This becomes apparent if we find that multiple NVSS/SUMSS sources coincide with the G4Jy source being inspected. We note such cases by updating a `confusion flag' to `1' (from a default value of `0') if the following criteria are satisfied: 
\begin{enumerate}
\item{An unrelated source is detected above 6\,$\sigma$ in NVSS/SUMSS.}
\item{The position of the peak NVSS/SUMSS emission for the unrelated source is within the 3-$\sigma$ GLEAM contour for the G4Jy source.}
\end{enumerate}

One reason we do not also consider TGSS for this is that we wish the confusion flag to be consistent (i.e. based upon the same spatial resolution) across the full sample, whereas TGSS is only available above Dec.~$=-53^{\circ}$. Furthermore, we have noticed artefacts in TGSS images surrounding bright sources (section~5.2.1 of Paper~I), and wish to avoid these being interpreted as real emission. Doing so would lead to the G4Jy source being incorrectly labelled as `confused'.

Of the 1,863 sources in the sample, 383 (21\%) have their confusion flag set to `1'.

\subsection{Host-galaxy identification flags}
\label{sec:identifyhost}

Assessing the radio morphology (Section~\ref{sec:morphology}) in conjunction with the distribution of mid-infrared (AllWISE) sources allows us to consider which galaxy is most-likely hosting the radio emission. For each G4Jy source we then assign one of the following four characters as the `host flag': 
\begin{itemize}
\item{`i' -- a host galaxy has been identified in the AllWISE catalogue, with the position and mid-infrared magnitudes (W1, W2, W3, W4) recorded as part of the G4Jy catalogue (Paper~I),}
\item{`u' -- it is unclear which AllWISE source is the most-likely host galaxy, due to the complexity of the radio morphology and/or the spatial distribution of mid-infrared sources (leading to ambiguity),}
\item{`m' -- identification of the host galaxy is limited by the mid-infrared data, with the relevant source either being too faint to be detected in AllWISE, or affected by bright mid-infrared emission nearby,}
\item{`n' -- no AllWISE source should be specified, given the type of radio emission involved.} 
\end{itemize}

The above assessment is done alongside consultation of the literature, with our considerations summarised in section~5.5.1 of Paper~I. In the next two sections (Sections~\ref{sec:variety} and \ref{sec:additionalchecks}) we present our findings for individual sources, having performed these further checks. Considering the full sample, 1,606 sources have host flag `i', 129 sources are labelled `u', 126 are labelled `m', and the remaining two sources are labelled `n'.

\section{A variety of radio sources}
\label{sec:variety}

A consequence of the high flux-density threshold used to define the G4Jy Sample is that the sample is dominated by powerful radio-galaxies. These come in a variety of shapes and sizes, as we will show in this section. However, the sample also contains sources where the bright radio emission is the result of star formation or merging halos, which we also highlight below.

\subsection{The Flame Nebula, a H\,{\it {\sc ii}} region}
\label{sec:nebula}

\begin{figure*}
\centering
\includegraphics[scale=2.0]{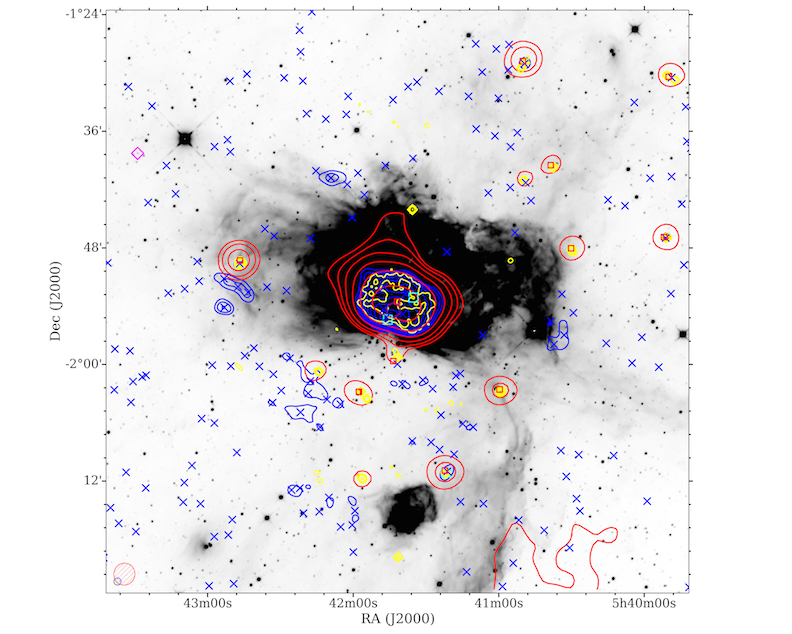}
\caption{An overlay (1\arcdeg\ across) for G4Jy~571, the Flame Nebula, centred on the component GLEAM~J054141$-$015331 (Section~\ref{sec:nebula}). Radio contours from TGSS (150\,MHz; yellow), GLEAM (170--231\,MHz; red), and NVSS (1.4\,GHz; blue), are overlaid on a mid-infrared image from {\it WISE} (3.4\,$\mu$m; inverted greyscale). For each set of contours, the lowest contour is at the 3\,$\sigma$ level (where $\sigma$ is the local rms), with the number of $\sigma$ doubling with each subsequent contour (i.e. 3, 6, 12\,$\sigma$, etc.). Also plotted, in the bottom left-hand corner, are ellipses to indicate the beam sizes for TGSS (yellow with `+' hatching), GLEAM (red with `/' hatching), and NVSS (blue with `\textbackslash' hatching). The centroid position is indicated by a purple hexagon, and also plotted are catalogue positions from TGSS (yellow diamonds), GLEAM (red squares), NVSS (blue crosses), AT20G (cyan squares), and 6dFGS (magenta diamonds).}
\label{FlameNebulaImage}
\end{figure*}

\begin{figure}
\centering
\includegraphics[scale=0.4]{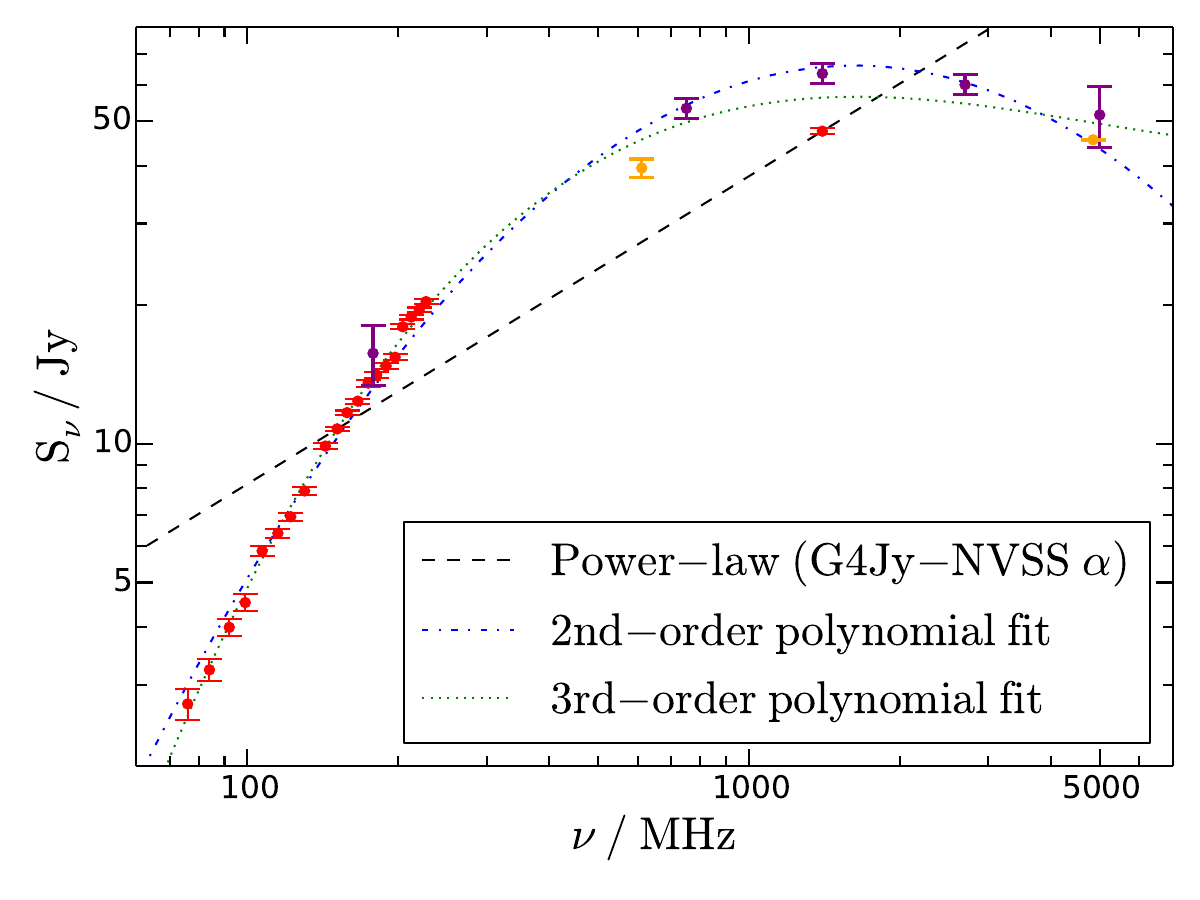}
\caption{The radio spectrum, from 72 to 5000\,MHz, for G4Jy~571 (the Flame Nebula; Section~\ref{sec:nebula}). Red datapoints represent measurements in the G4Jy catalogue, from GLEAM and NVSS (over 72--231\,MHz and at 1400\,MHz, respectively). Purple datapoints represent measurements from \citet{Kellermann1969}, at 178, 750, 1400, 2695, and 5000\,MHz. Orange datapoints represent measurements by \citet{Durdin2000} at 611\,MHz, and  \citet{Griffith1995} at 4850\,MHz. We also plot a power-law (black, dashed line) that uses the spectral index calculated between 151 and 1400\,MHz (G4Jy\_NVSS\_alpha in the G4Jy catalogue), in addition to 2nd-order (blue, dash-dotted line) and 3rd-order (green, dotted line) polynomial fits to the data in log-log space.}
\label{FlameNebulaSED}
\end{figure}

One of the few non-AGN sources in the G4Jy Sample is the Flame Nebula ({\bf G4Jy~571}; GLEAM~J054141$-$015331; Figure~\ref{FlameNebulaImage}). This is a H\,{\sc ii} region found in the constellation of Orion, and is also referred to as Orion B, NGC~2024, PKS~B0539$-$019, and 3C~147.1. It is at a distance of $363\,\pm\,75$\,pc \citep{Bik2003}, and we present the radio spectrum for this star-forming region in Figure~\ref{FlameNebulaSED}. Below $\sim$1\,GHz the thermal emission for this source becomes optically thick, hence the turnover in the spectrum. [The reason we still plot a power-law function is to demonstrate that spectral indices associated with such descriptions (section~6.6 of Paper~I) need to be used with due consideration of spectral curvature.] Since the associated mid-infrared emission is nearby and extended, we do not specify an AllWISE position for this source. 

\subsection{Star-forming galaxies}
\label{sec:SFGs}

\begin{figure*}
\centering
\subfigure[G4Jy~86]{
	\includegraphics[scale=1.1]{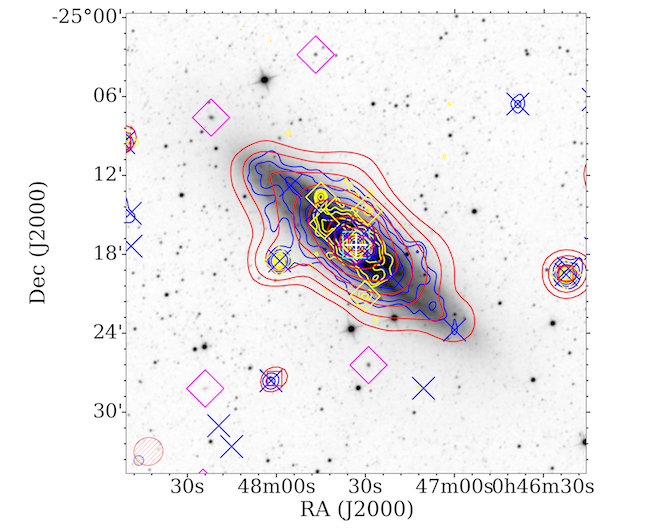}
	}
\subfigure[G4Jy~1081]{
	\includegraphics[scale=1.1]{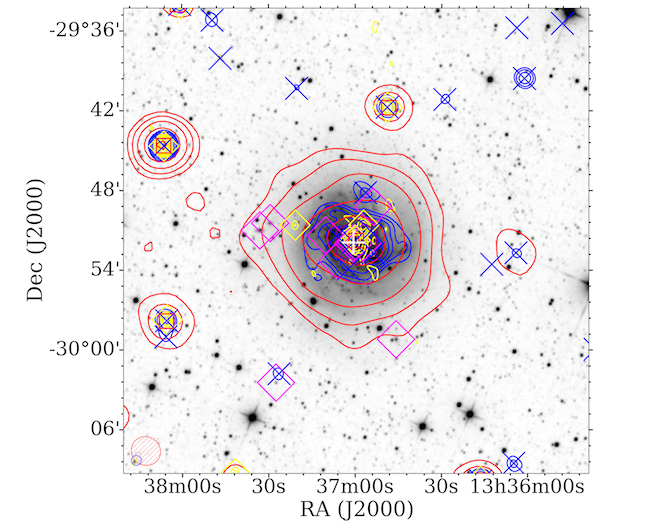}
	}
\subfigure[G4Jy~150 and G4Jy~151]{
	\includegraphics[scale=1.1]{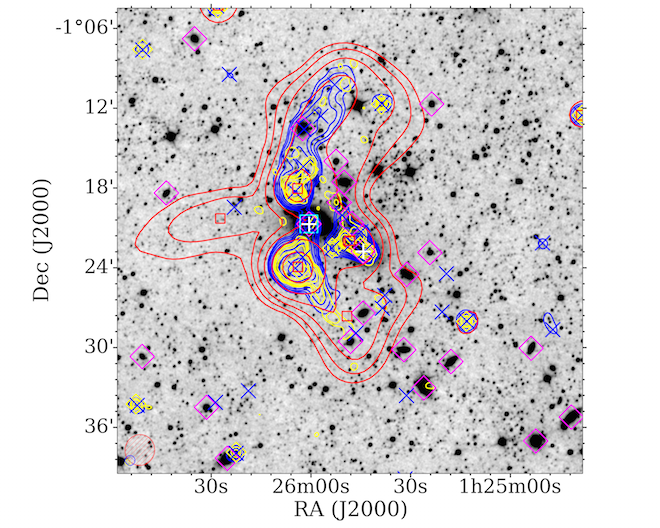}
	}
\subfigure[G4Jy~1110]{
	\includegraphics[scale=1.1]{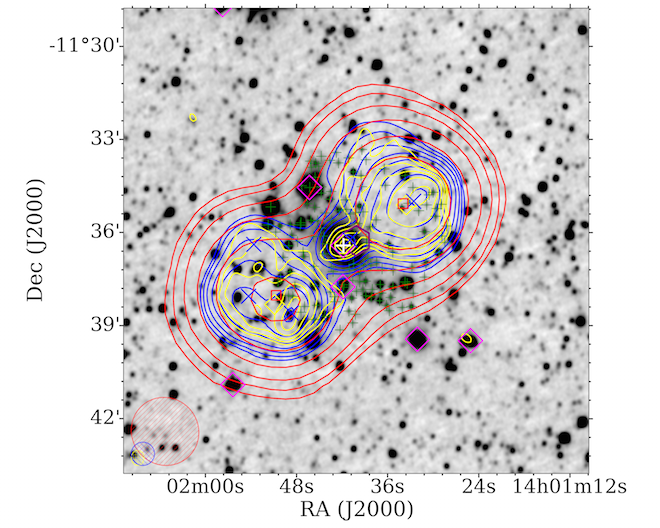}
	}
\caption{Overlays for (a) the Sculptor Galaxy, G4Jy~86, (b) the Southern Pinwheel Galaxy, G4Jy~1081, (c) 3C~40, G4Jy~150 (white plus sign towards the west) and G4Jy~151 (white plus sign towards the east), and (d) G4Jy~1110 (Sections~\ref{sec:SFGs}--\ref{sec:minkowski} and appendix~A of G4Jy~Paper~III; White et al., in prep.). The datasets, contours, symbols, and beams are the same as those used for Figure~\ref{FlameNebulaImage}. In addition, white plus signs indicate the host-galaxy positions for G4Jy sources, and we use a logarithmic scale for the inverted, mid-infrared images in the first two panels. For panels (c) and (d), the usual inverted, linear greyscale is used. \label{varietyoverlays} }
\end{figure*}

{\bf G4Jy~86} (GLEAM~J004733$-$251710) is NGC~253, also known as the Sculptor Galaxy (Figure~\ref{varietyoverlays}a). This is an intermediate spiral galaxy, whose proximity has enabled numerous detailed studies of its intense star-formation \citep[e.g.][]{Rieke1980,Ulvestad1997,Lenc2006}. Recently, \citet{Kapinska2017} studied the broadband radio spectrum for this source (from 76\,MHz to 11\,GHz), where the emission is described as a central starburst region in combination with an extended radio halo. TGSS misses some of the total flux-density for nearby sources, but thanks to the tight constraints provided by GLEAM measurements, \citet{Kapinska2017} find that the central component is best-modelled as synchrotron plasma that exhibits free-free absorption at frequencies below $\sim$230\,MHz.

Another star-forming, spiral galaxy in the G4Jy Sample is {\bf G4Jy~1081} (GLEAM~J133659$-$295147). This is more commonly known as M83 (NGC~5236), and is nicknamed the Southern Pinwheel Galaxy (Figure~\ref{varietyoverlays}b). Like G4Jy~86, this nearby starburst is well-studied at multiple wavelengths. For example, sub-arcsecond resolution images in the near-infrared \citep{Gallais1991} revealed that the double nucleus at the centre of the galaxy consists of an arc of star clusters offset from the old stellar nucleus. At radio wavelengths, \citet{Heald2016} present the extent of neutral hydrogen surrounding the galaxy, as well as its large-scale magnetic field.

\subsection{Radio emission near Minkowski's object}
\label{sec:minkowski}

`Minkowski's object', originally identified by  \citet{Minkowski1958}, is a well-known starburst whose star formation appears to have been triggered by a radio jet emanating from NGC~541 \citep{vanBreugel1985,Croft2006}. NGC~541 is listed in the G4Jy Sample as {\bf G4Jy~150} (GLEAM~J012548$-$012157), and along with the dumb-bell galaxy NGC~545/NGC~547, resides in cluster Abell~194. NGC~547 is the host of {\bf G4Jy~151} (GLEAM~J012603$-$012356 and GLEAM~J012604$-$011802), as confirmed by archival VLA data at 1.5\,GHz \citep{Jetha2006,Sakelliou2008}. Together, G4Jy~150 and G4Jy~151 make up the (confused) radio source known as 3C~40 (Figure~\ref{varietyoverlays}c). Being at $z=0.018$, the 26$'$ angular extent of G4Jy~151 corresponds to a projected, linear size of 540\,kpc. Interestingly, the GLEAM contours (for GLEAM~J012627$-$012017) suggest the presence of an `extra' lobe in the system, towards the east and possibly associated with NGC~541 or NGC~545. We note that this feature has been observed previously by \citet{Sakelliou2008} and \citet{Govoni2017}, but no conclusive interpretation has been made.

\subsection{`X-' and `S-'/`Z-shaped' sources}
\label{sec:XandSorZshaped}

Rare sources picked up by the GLEAM Survey include those showing possible intermittent jet activity. In some cases the orientation of the jet may have changed between, or during, episodes of jet production. For the former scenario, the radio source may exhibit an `X-shaped' appearance \citep[e.g.][]{Capetti2002,Saripalli2009}. By using the spectral information provided by the MWA, in combination with data of higher spatial resolution at low frequencies, it is possible to calculate the ages of the two sets of lobes, and thereby estimate the duty cycle of the AGN. However, an alternative explanation for X-shaped morphology is that the source is in an ongoing group or cluster merger, with hydrodynamical effects and the surrounding medium influencing the radio structure \citep[e.g.][]{Hardcastle2019}. Whatever the origin or subsequent processing of the emission, there are eight G4Jy sources with X-shaped morphology (Section~\ref{sec:Xshapedsources}), six of which are shown in Figure~\ref{Xshape}.

Meanwhile, it has been suggested that a less-abrupt change in the jet axis is what leads to `S-shaped' or `Z-shaped' radio-sources (Figure~\ref{Zshape}a). We find six such sources in the G4Jy Sample, which we detail in Section~\ref{sec:Zshapedsources}. For these cases, we see (clearer) evidence of precession of the radio jets, which is thought to be caused by two supermassive black-holes orbiting one another \citep[e.g.][]{Hunstead1984,Merritt2002}. 

\subsubsection{X-shaped sources}
\label{sec:Xshapedsources}

\begin{figure*}%
\centering
\subfigure[A 10$'$ overlay of G4Jy~18]{
	\includegraphics[scale=1.1]{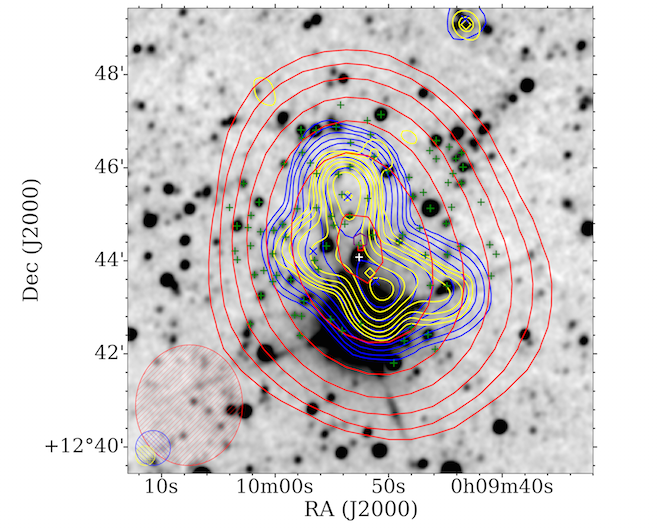} 
	} 
\subfigure[A 10$'$ overlay of G4Jy~105]{
	\includegraphics[scale=1.1]{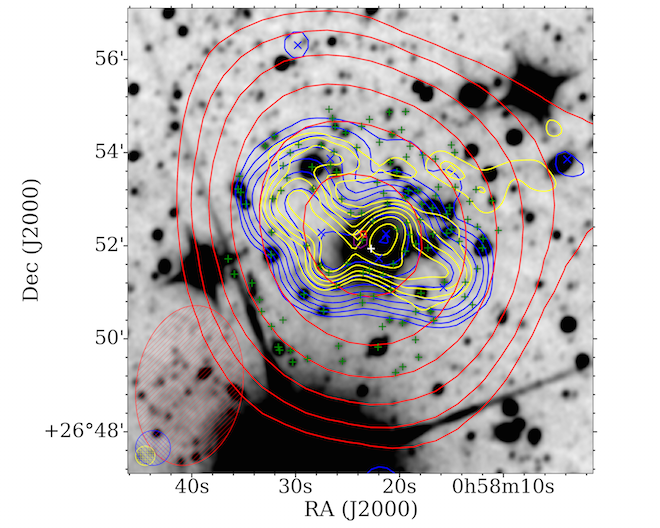} 
	} 
\subfigure[A 10$'$ overlay of G4Jy~683]{
	\includegraphics[scale=1.1]{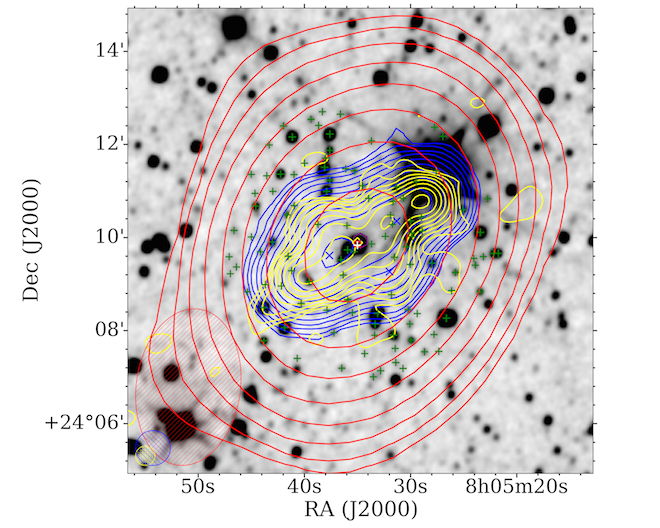} 
	} 
\subfigure[A 10$'$ overlay of G4Jy~1233]{
	\includegraphics[scale=1.1]{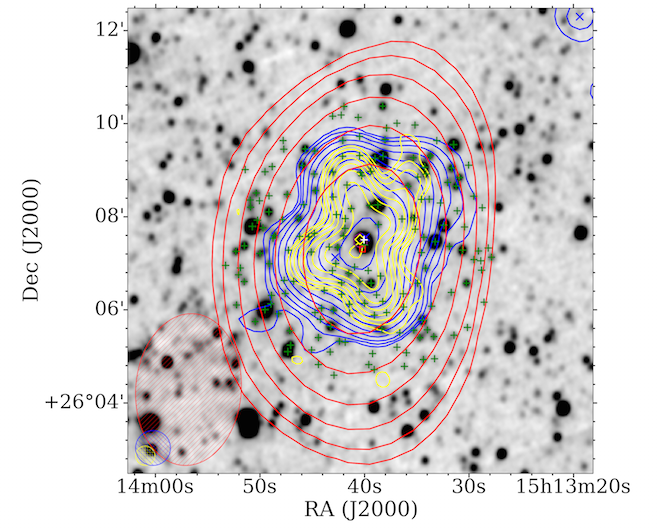} 
	} 
\subfigure[A 1\arcdeg\ overlay of G4Jy~1613]{
	\includegraphics[scale=1.1]{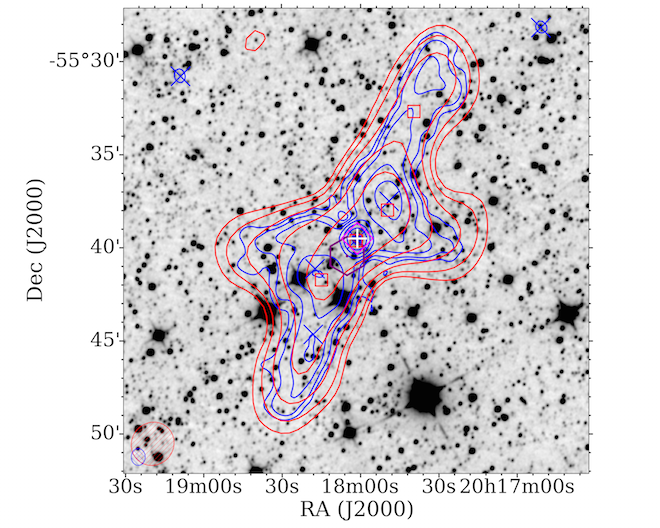} 
	} 
\subfigure[A 10$'$ overlay of G4Jy~1846]{
	\includegraphics[scale=1.1]{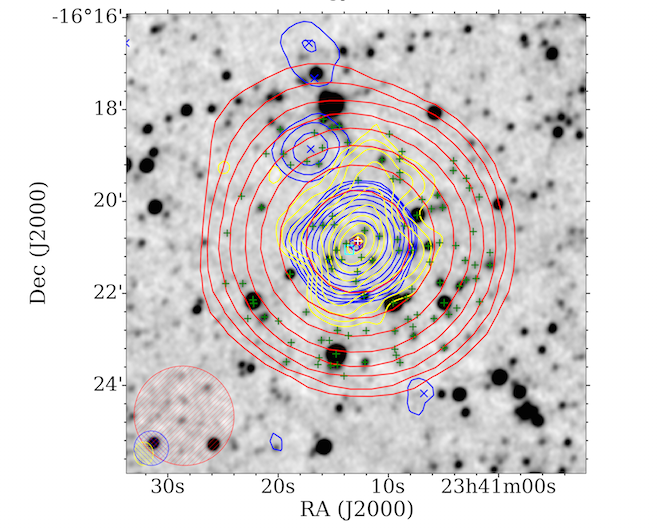} 
	} 
\caption{Six of the eight X-shaped radio-sources in the G4Jy Sample (Section~\ref{sec:Xshapedsources}). The datasets, contours, symbols, and beams are the same as those used for Figure~\ref{FlameNebulaImage}, but where blue contours, crosses, and ellipses correspond to NVSS {\it or} SUMSS. Host galaxies are highlighted with a white plus, and for the 10$'$ overlays, all AllWISE positions within 3$'$ of the centroid are also shown (green plus signs). \label{Xshape} }
\end{figure*}

The first X-shaped source in the G4Jy Sample, {\bf G4Jy~18} (GLEAM~J000952+124416; Figure~\ref{Xshape}a), also appears in the 3CRR catalogue and is known as 4C~+12.03. Observations of the radio core \citep{Leahy1991,Kuzmicz2017} allow us to confirm that we have correctly (manually) identified the host galaxy.

NGC~326 (4C~+26.03) is listed as {\bf G4Jy~105} (GLEAM~J005823+265214; Figure~\ref{Xshape}b). This source has been observed by \citet{Murgia2001} at multiple frequencies using the VLA, and their figure~9 clearly shows that the radio core is associated with the northern member of the dumb-bell galaxy. We dismiss the AllWISE position closest to the centroid position, and confirm that the newly-chosen host-galaxy position is in agreement with the core position (subject to the resolution of AllWISE). A new LOw-Frequency ARray (LOFAR) image of this source reveals that the northernmost tip of the `X' extends into a long plume \citep{Hardcastle2019}, a hint of which is seen in the TGSS contours. 

{\bf G4Jy~683} (GLEAM~J080535+240951) is also in the 3CRR catalogue, as 3C~192 (Figure~\ref{Xshape}c). The density of several mid-infrared sources close to the centroid again prompts us to check against observations of the core  \citep{Baum1988,Leahy1997}. Doing so confirms that the original candidate -- the AllWISE source closest to the centroid -- is correct.

Also overlapping with 3CRR is {\bf G4Jy~1233} (GLEAM~J151340+260718; 3C~315). This source has two AllWISE positions close to the centroid (Figure~\ref{Xshape}d), but the clear image of the core presented by \citet{Leahy1986} allows the host galaxy to be confirmed.

{\bf G4Jy~1581} (GLEAM~J195215+023032) can be found in the Revised Third Cambridge Catalogue \citep[3CR;][]{Bennett1962,Spinrad1985} as 3C~403. Its X-shaped morphology is evident in the TGSS contours, and a VLA image by \citet{Black1992} confirms that the AllWISE source closest to the centroid is the host galaxy.

{\bf G4Jy~1613} (B2014$-$558/J2018$-$556) is perhaps the best-studied X-shaped source in the southern hemisphere, thanks to its large spatial extent ($\sim$20$'$ along the longer-lobe axis\footnote{NVSS and SUMSS were the catalogues used to estimate the angular sizes of sources in the G4Jy catalogue (see section 6.3.1 of Paper~I). Therefore, readers should bear in mind that the angular size provided may be an underestimate for extended sources such as G4Jy~1613, due to NVSS or SUMSS components not being distributed along the full extent of the radio emission.}). Its low-frequency emission is characterised by four GLEAM components (GLEAM~J201739$-$553242, GLEAM~J201749$-$553800, GLEAM~J201801$-$553938, and GLEAM~J201814$-$554145) and the host galaxy is easily identified as g2018013$-$553932 in 6dFGS (Figure~\ref{Xshape}e). This provides the redshift ($z=0.0606$), from which we estimate the physical size of the radio source at 1.4\,Mpc. This means that G4Jy~1613 also falls under the category of `giant radio-galaxy' (see Section~\ref{sec:GRGs}).

Whilst {\bf G4Jy~1846} (GLEAM~J234112$-$162052; PKS~B2338$-$166) appears to show X-shaped contours in TGSS (Figure~\ref{Xshape}f), we are unable to find another radio image in the literature to support this. However, the coincidence of three radio positions suggests that the chosen host-galaxy is robust, with the nearby AT20G detection likely being the result of hotspot emission.

Another X-shaped source in the sample is {\bf G4Jy~1122} (GLEAM~J140649$-$015417). Being unresolved in TGSS, its morphology is only revealed by the 5$''$ resolution of FIRST (Faint Images of the Radio Sky at Twenty-Centimeters; \citealt{White1997}). Hence, we expect follow-up, high-resolution observations of the remainder of the G4Jy Sample to uncover additional sources such as these.

\subsubsection{S-/Z-shaped sources}
\label{sec:Zshapedsources}

\begin{figure*}
\centering
\subfigure[G4Jy~241]{
	\includegraphics[scale=1.0]{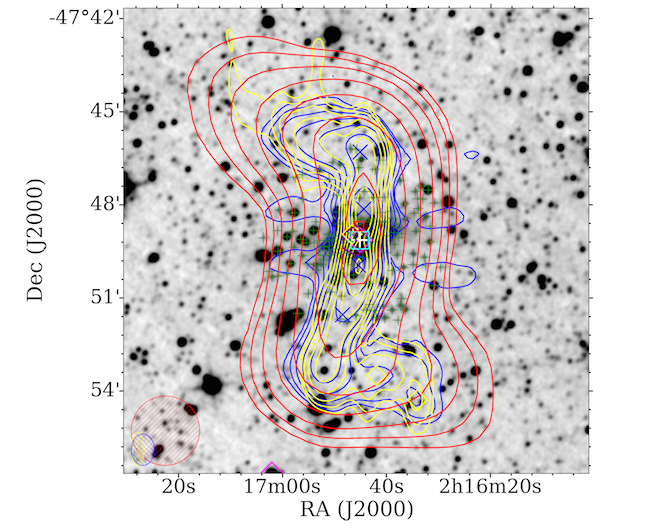}
	}
\subfigure[G4Jy~480]{
	\includegraphics[scale=1.0]{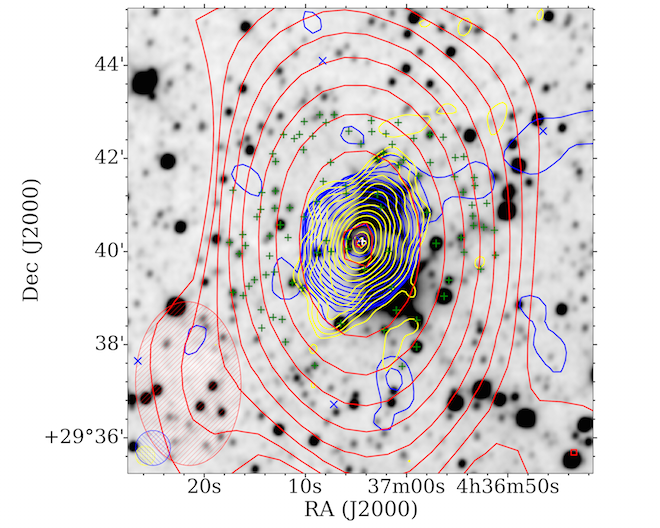}
	}
\caption{Two of the six S-/Z-shaped radio-sources in the G4Jy Sample (Section~\ref{sec:Zshapedsources}), where the Z-shaped morphology of G4Jy~480 is revealed at 1.7\,GHz (see \citealt{Cox1991}). The datasets, contours, symbols, and beams are the same as those used for Figure~\ref{FlameNebulaImage}, but where blue contours, crosses, and ellipses correspond to NVSS {\it or} SUMSS. In addition, positions from AllWISE are indicated by green plus signs, with that corresponding to the host galaxy highlighted in white. \label{Zshape}}
\end{figure*}

{\bf G4Jy~241} (GLEAM~J021645$-$474842) is the brightest cluster-galaxy (BCG) of cluster AS~239 \citep{Abell1989}. The radio morphology in the inner region, alongside the coincidence of 6dFGS and AT20G detections (Figure~\ref{Zshape}a), allow us to confidently identify the host galaxy (ESO~198-1). At $z=0.064$ (g0216451-474909), the angular size of the radio emission ($\sim8'$) corresponds to a physical size of 590\,kpc. 

The S-shaped radio source, {\bf G4Jy~447} (GLEAM~J042220+140742, GLEAM~J042233+140733), became a member of the G4Jy Sample following an internal match of the EGC (see section~7.2.1 of Paper~I). It is also known as 4C~+14.12 (PKS~B0419+14) and we select the bright mid-infrared source, coinciding with the narrowing of the TGSS contours (figure~8d of Paper~I), as the host galaxy.   

{\bf G4Jy~543} (GLEAM~J052522$-$324121 and GLEAM~J052531$-$324357) is PKS~B0523$-$32 in cluster AS 527 \citep{Abell1989}. It was added to the G4Jy Sample following a comparison with the sample of \citet{Jones1992} [see section~7.1.1 of Paper~I], and we follow their host-galaxy identification. This is on account of its 6dFGS detection (g0525272$-$324216) being positioned at the `pinch' of the TGSS contours (figure~4b of Paper~I). G4Jy~543 is $\sim7'$ across, which corresponds to a physical size of 610\,kpc at $z=0.077$.  

For {\bf G4Jy~1523} (GLEAM~J191757$-$243917), a nearby star (J19175653$-$2438509) leads to obscuration in the mid-infrared image. This means that we are unable to identify the host galaxy in AllWISE, and so we set the host flag to `m' in the G4Jy catalogue.

Of course, radio images of better resolution would allow additional S-/Z-shaped sources to be identified. For example, {\bf G4Jy~480} (GLEAM~J043704+294009) is 3C~123 in 3CRR. It shows Z-shaped symmetry at 1.7\,GHz (observations by Laing, published by \citealt{Cox1991}), but only has rhombus-shaped contours at the 25$''$ resolution provided by TGSS (Figure~\ref{Zshape}b). 

Similarly, {\bf G4Jy~1802} (GLEAM~J230303$-$184129) shows extended emission in NVSS and TGSS, but its S-shaped morphology is revealed in 1.5-GHz images by \citet{Hunstead1984}. They note an optically-faint companion galaxy at the same redshift as the host of G4Jy~1802 ($z = 0.129$), but its separation of 6.8$''$ on the sky (and the size of the host in the W1 image) means that it is not distinguished in AllWISE. However, our host-galaxy identification (AllWISE~J230302.97$-$184125.8) is reliable, with its position coinciding with both a detection in AT20G and in 6dFGS.

\subsection{A cluster relic and a halo}
\label{sec:clusterrelics}

\begin{figure*}
\centering
\includegraphics[scale=2.0]{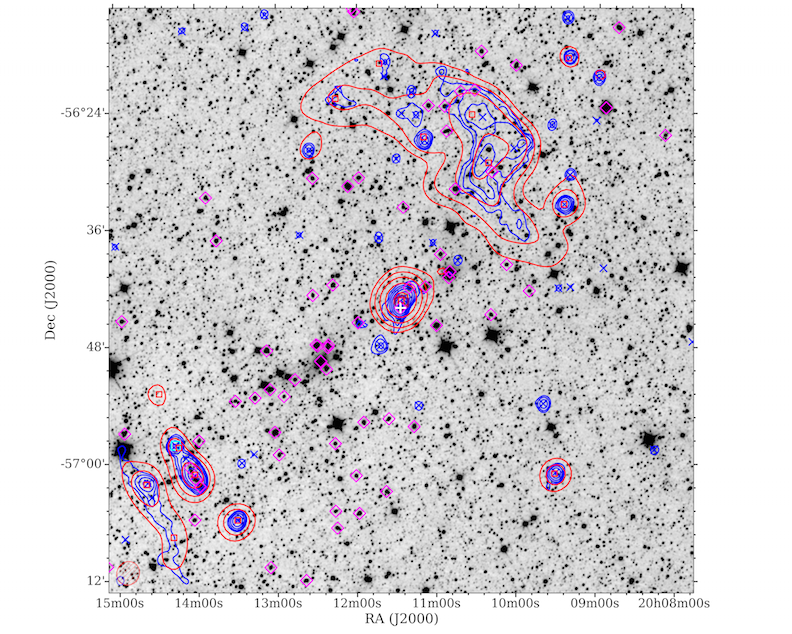}
\caption{A 1\arcdeg\ overlay of cluster Abell~3667, where G4Jy~1605 is the cluster relic in the north-west (Section~\ref{sec:clusterrelics}), and G4Jy~1606 is the head-tail radio-galaxy at the centre (Section~\ref{subsec:headtails}). The host galaxy for G4Jy~1606 is highlighted with a white plus sign. Radio contours from GLEAM (170--231\,MHz; red) and SUMSS (843\,MHz; blue) are overlaid on a mid-infrared image from {\it WISE} (3.4\,$\mu$m; inverted greyscale). For each set of contours, the lowest contour is at the 3\,$\sigma$ level (where $\sigma$ is the local rms), with the number of $\sigma$ doubling with each subsequent contour (i.e. 3, 6, 12\,$\sigma$, etc.). Also plotted, in the bottom left-hand corner, are ellipses to indicate the beam sizes for GLEAM (red with `/' hatching) and SUMSS (blue with `\textbackslash' hatching). The centroid position for G4Jy~1606 is indicated by a purple hexagon, and also plotted are catalogue positions from GLEAM (red squares), SUMSS (blue crosses), AT20G (cyan squares), and 6dFGS (magenta diamonds).
\label{clusterrelic}}
\end{figure*}

The G4Jy Sample contains several examples of radio emission associated with galaxy clusters, including the extended radio-emission from cluster Abell~3667 \citep{Rottgering1997,Hindson2014}. This emission is thought to be a consequence of two halos merging \citep[e.g.][]{Ensslin1998}, with shock waves having propagated outwards, resulting in a radio relic on opposite sides of the cluster (Figure~\ref{clusterrelic}). As five GLEAM components are associated with each other, we update the centroid position for the northern relic and list this source as {\bf G4Jy~1605}. [The southern relic is composed of GLEAM~J201418$-$570734 and GLEAM~J201438$-$570209, whose summed flux-density at 151\,MHz is below 4\,Jy. As such, this relic is not included in the G4Jy Sample.] Since it is inappropriate to cross-identify the relic with a mid-infrared source, we set the host flag to `n' and label the morphology `complex'. With cluster members at an average redshift of $z = 0.055$ \citep{JohnstonHollitt2008,Owers2009}, the 1\arcdeg\ separation of the two relics equates to a physical extent of 3.9\,Mpc. 

\begin{figure*}
\centering
\subfigure[G4Jy~77]{
	\includegraphics[scale=1.1]{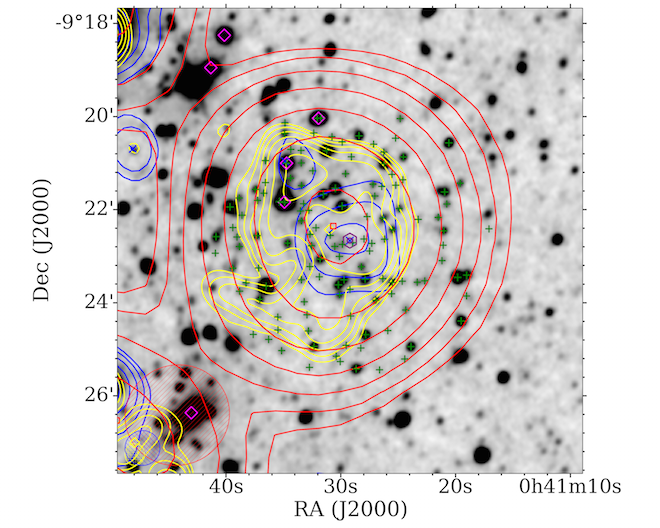} 
	}
\subfigure[G4Jy~1550]{
	\includegraphics[scale=1.1]{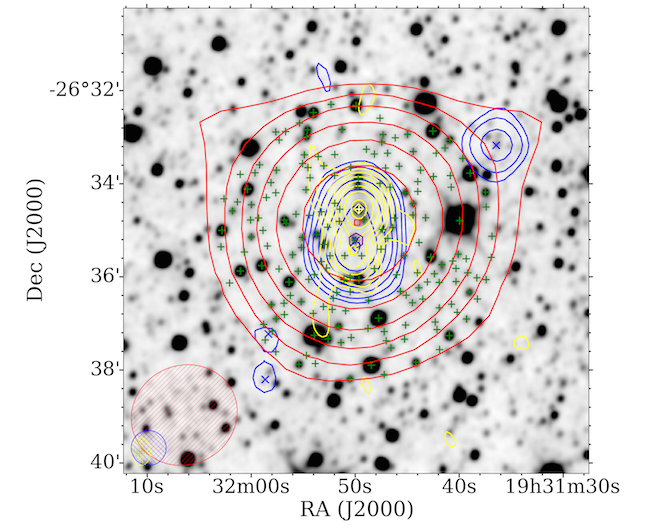} 
	}
\caption{Two 10$'$ overlays, of (a) G4Jy~77 in cluster Abell~85 (Section~\ref{sec:clusterrelics}), and (b) the lensing cluster MACS~J1931.8$-$2634, which contains G4Jy~1550 (Section~\ref{sec:lensingcluster}). The datasets, contours, symbols, and beams are the same as those used for Figure~\ref{FlameNebulaImage}, with positions from AllWISE indicated by green plus signs. For G4Jy~1550, the host galaxy is highlighted in white. \label{halolensingcluster}}
\end{figure*}

Our attention is also drawn to the unusual morphology, and very steep-spectrum emission ($\alpha = -2.64 \pm 0.04$, measured between 151 and 1400\,MHz), of {\bf G4Jy~77} (GLEAM~J004130$-$092221; Figure~\ref{halolensingcluster}a). It is B0038$-$096 in cluster Abell~85, with \citet{Bagchi1998} interpreting it as a radio {\it halo}. It is unclear whether this plasma was energised during a cluster merger, or by a radio galaxy that has long-ago `switched off', so we leave the host flag as `u' (rather than `n'). For the morphology, we label this source `complex'. 

\subsection{A lensing cluster}
\label{sec:lensingcluster}

The TGSS and NVSS contours (and positions) for GLEAM~J193149$-$263450 suggest that there are two unrelated radio-sources close together, with different spectral indices (Figure~\ref{halolensingcluster}b). The GLEAM integrated flux-density at 151\,MHz is 6.16\,Jy, with TGSS indicating that the northern source (i.e. the northern TGSS component) accounts for 81\% of the low-frequency emission. Its flux density is greater than 4\,Jy and so GLEAM~J193149$-$263450 remains in the G4Jy Sample (with its confusion flag already set to `1' on account of the NVSS point-source towards the north-west). We refer to this northern source as {\bf G4Jy~1550}, which has a spectral index (between 151\,MHz and 1.4\,GHz) of $\sim-2.3$. The southern source (i.e. the southern TGSS component) has a spectral index of $\sim-1.9$, and is suggested to be a mini-halo [although \citet{Giacintucci2014} caution that such a classification is very uncertain]. These sources belong to the known lensing cluster MACS~J1931.8$-$2634, and G4Jy~1550 (the BCG) is found to have equal AGN- and starburst-contributions to its ultra-luminous far-infrared emission \citep{Santos2016}. As we do not have sufficient resolution at $\sim1$\,GHz to update the centroid position -- and we wish to maintain consistency across the sample -- it remains determined by the single NVSS component that we consider as associated with G4Jy~1550. The 1.4\,GHz flux-density quoted in the G4Jy catalogue is therefore an overestimate. As for the host galaxy of G4Jy~1550, this is the AllWISE source coincident with the northern TGSS position.

\subsection{Bent- and head-tail radio-galaxies}
\label{sec:tails}

Other radio sources connected with clusters are radio galaxies with `bent' lobes or `twin tails'. These are also referred to as WAT (wide-angle tail) and NAT (narrow-angle tail) radio-galaxies \citep[see][and references therein]{Miley1980}. This morphology is indicative of the radio galaxy falling into a cluster, with its radio lobes getting pushed backwards by ram pressure from the surrounding medium. As a result, a `double' may no longer be axisymmetric, and so we need to search for the host galaxy near the apex of the radio emission. Furthermore, the projection of the radio lobes (or `tails') with respect to the line-of-sight may give the appearance that they are overlapping. This means that the host galaxy is now at the `head' of the radio emission (hence the term `head-tail' galaxy; e.g. \citealt[][]{Miley1973}), rather than being located between two distinct radio-lobes. As such, we describe the morphology of head-tail galaxies as `complex'. We find a total of 41 bent/head-tail radio-galaxies in the G4Jy Sample, and list them in Tables~\ref{listofbenttails} and \ref{listofheadtails}. 

\subsubsection{Bent-tail radio-galaxies}
\label{subsec:benttails}

\begin{table}
\centering 
\caption{A list of G4Jy sources that have bent-tail morphology, which we label as `double' or `triple' in the G4Jy catalogue (see Sections~\ref{sec:morphology} and \ref{subsec:benttails}). Such interpretation of the morphology informs our identification of the host galaxy in the AllWISE image.}
\begin{tabular}{@{}rcc@{}}
\hline
Source  & GLEAM component(s) & Morphology  \\
\hline
G4Jy~47  & GLEAM~J002530$-$330336  & double \\   
G4Jy~315 & GLEAM~J025738+060352 & \\
 & GLEAM~J025748+060201 & double \\
 G4Jy~366 & GLEAM~J033401$-$385840 & \\ & GLEAM~J033416$-$390129 & double \\
 G4Jy~367  & GLEAM~J033414$-$011121 & double \\
 G4Jy~462  & GLEAM~J042839$-$535020 & \\ & GLEAM~J042907$-$534919 & triple \\
G4Jy~607 & GLEAM~J062140$-$524109 & triple \\
G4Jy~637 & GLEAM~J070130+231313 & triple \\   
G4Jy~651 & GLEAM~J071706$-$362140 & double \\
G4Jy~693 & GLEAM~J081630$-$703925 & triple \\
G4Jy~949  & GLEAM~J114507+193718 & triple \\
G4Jy~1004 & GLEAM~J122950+114015 & double \\  
G4Jy~1011 & GLEAM~J123629+163201 & double \\
G4Jy~1034 & GLEAM~J125437$-$123333 & double \\  
G4Jy~1060 & GLEAM~J131616+070219 & double \\
G4Jy~1067  &    GLEAM~J132606$-$272641 & \\
       &    GLEAM~J132616$-$272632 & double \\  
G4Jy~1094 & GLEAM~J134855$-$252700 & double \\  
G4Jy~1173  & GLEAM~J142955+072134 & \\
  & GLEAM~J143002+071505 & double \\
G4Jy~1245 & GLEAM~J152007$-$283411 & double \\    
G4Jy~1389  & GLEAM~J170752$-$222543 & double \\   
G4Jy~1496 & GLEAM~J183626+193946 & \\
 & GLEAM~J183640+194318 & \\
 & GLEAM~J183649+194105 & triple \\  
G4Jy~1544 & GLEAM~J192819$-$293157 & triple \\ 
G4Jy~1704  & GLEAM~J213356$-$533509 & \\
 & GLEAM~J213418$-$533514  & double \\
G4Jy~1705  & GLEAM~J213415$-$533736 & \\
 & GLEAM~J213422$-$533756  & double \\
\hline
\label{listofbenttails}
\end{tabular}
\end{table}

\begin{figure*}%
\centering
\subfigure[G4Jy~315]{
	\includegraphics[scale=1.1]{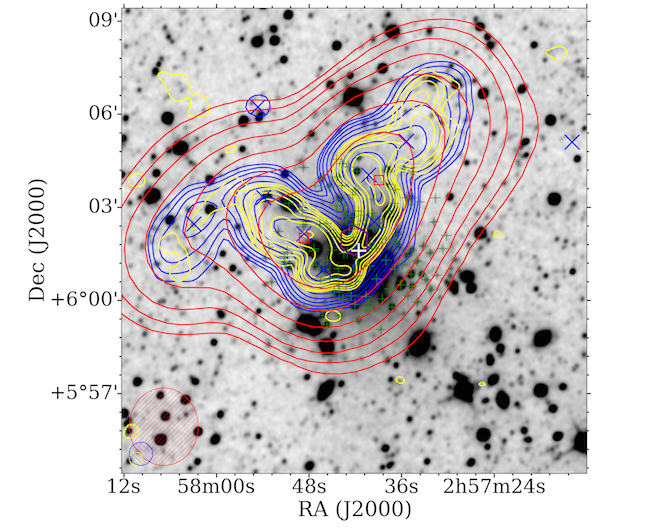} 
	} 
\subfigure[G4Jy~367]{
	\includegraphics[scale=1.1]{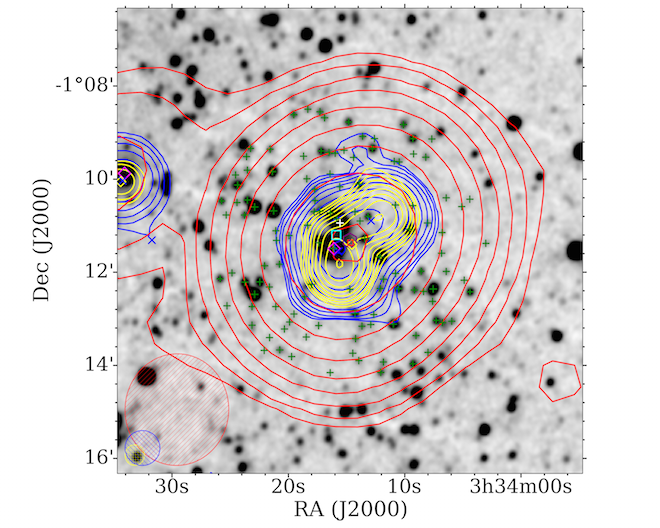} 
	} 
\subfigure[G4Jy~637]{
	\includegraphics[scale=1.1]{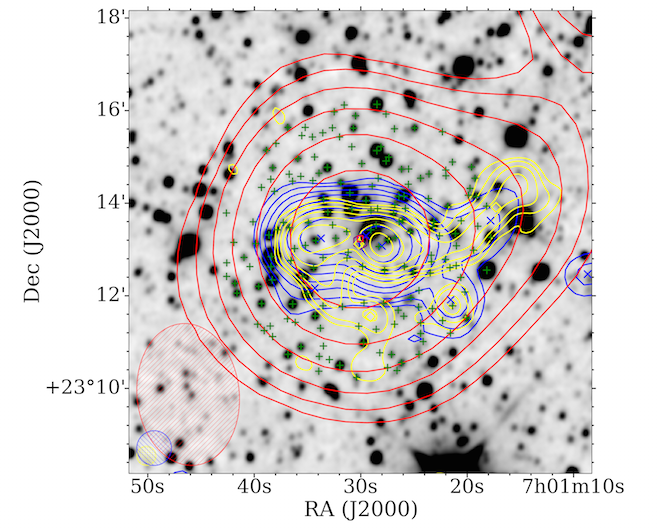} 
	} 
\subfigure[G4Jy~949]{
	\includegraphics[scale=1.1]{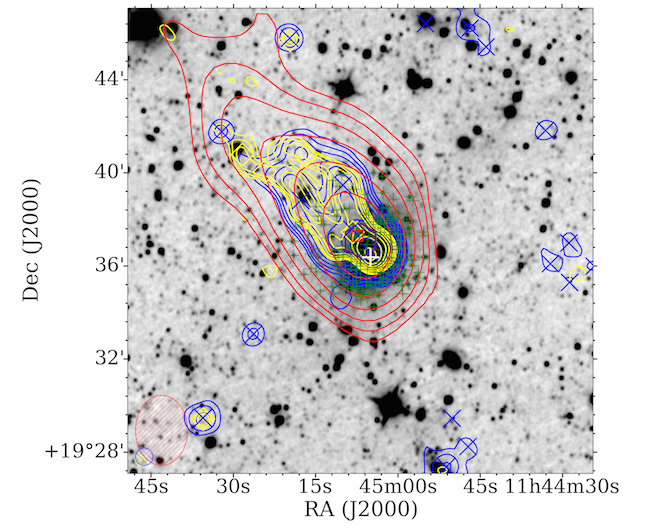} 
	} 
\subfigure[G4Jy~1094]{
	\includegraphics[scale=1.1]{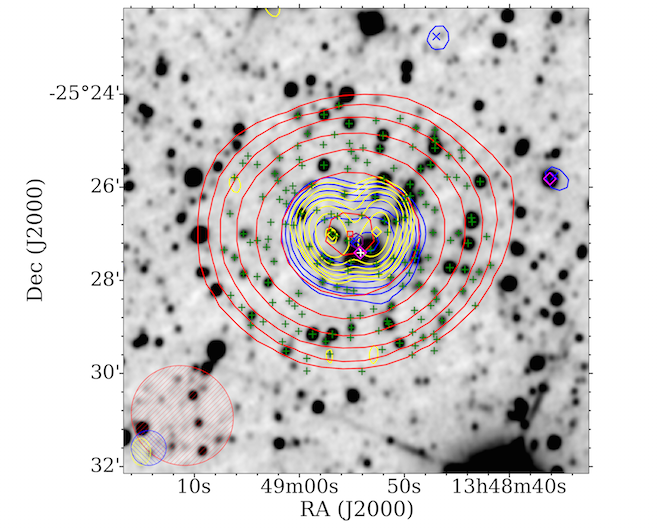} 
	} 
\subfigure[G4Jy~1245]{
	\includegraphics[scale=1.1]{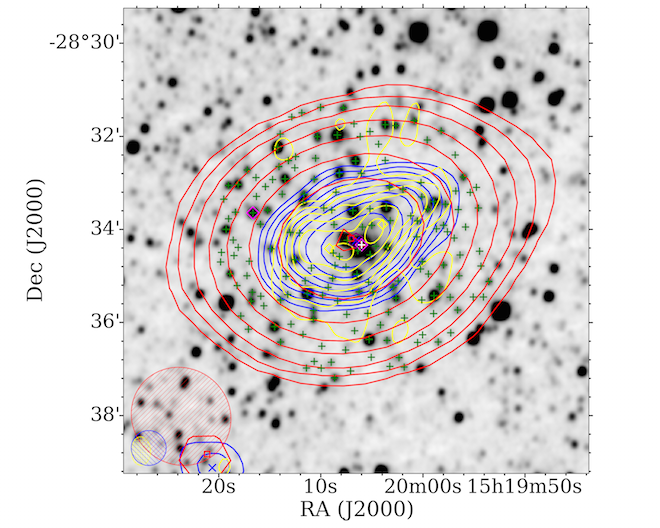} 
	}
\caption{Examples of bent-tail radio-galaxies in the G4Jy Sample (Section~\ref{subsec:benttails}). The datasets, contours, symbols, and beams are the same as those used for Figure~\ref{FlameNebulaImage}, but where blue contours, crosses, and ellipses correspond to NVSS {\it or} SUMSS. Host galaxies are highlighted with a white plus, and all AllWISE positions within 3$'$ of the centroid are also shown (green plus signs). \label{benttailoverlays} }
\end{figure*}

{\bf G4Jy~315} (GLEAM~J025738+060352 and GLEAM~J025748+060201) is a particularly interesting source, with checks against the literature showing that this is 3C~75 (NGC~1128) in cluster Abell 400. \citet{Owen1985} present 1.4$''$ resolution images at 4.9\,GHz that reveal that the radio emission is produced by {\it two} WAT radio-galaxies. Not having the resolution to disentangle these two sources at low frequencies (Figure~\ref{benttailoverlays}a), we proceed with updating the confusion flag to `1'. Also, we select the AllWISE position that is consistent with the northern radio-core -- the brighter of the two nuclei \citep{Liuzzo2010} -- following existing convention.


Our host-galaxy identification for {\bf G4Jy~367} (GLEAM~J033414$-$011121; 3C~89) is confirmed by a clear radio-core, and `triple' morphology, in FIRST. Therefore, somewhat unusually, the AT20G position (see Figure~\ref{benttailoverlays}b) possibly indicates where the {\it inner} part of the southward jet is colliding into the dense medium of cluster RXJ0334.2-0111 \citep{Dasadia2016}.

{\bf G4Jy~462} (GLEAM~J042839$-$535020 and GLEAM~J042907$-$534919) is IC~2082 (B0427$-$539) in cluster AS~463 \citep{Abell1989}. We agree with the host-galaxy identification of \citet{Jones1992} but interpret this source as a WAT radio-galaxy rather than a head-tail source, due to the extended SUMSS emission seen towards the north-east.

{\bf G4Jy~637} (GLEAM~J070130+231313) has a WAT morphology, as shown by the trailing TGSS contours (Figure~\ref{benttailoverlays}c). This source appears in the literature as 4C~+23.18 but we cannot find a high-resolution image to help distinguish between host-galaxy candidates. We therefore leave the host flag as `u'.


{\bf G4Jy~949} (GLEAM~J114507+193718) is 3C~264 (NGC~3862) in the 3CRR sample. In addition to {\it Hubble Space Telescope} images of the optical/radio jet, \citet{Baum1997} present MERLIN observations that allow us to select the correct AllWISE position for this NAT radio-galaxy (Figure~\ref{benttailoverlays}d).

The WAT radio-galaxy, {\bf G4Jy~1004} (GLEAM~J122950+114015), is in cluster Abell~1552. Due to sidelobes caused by M87 (Virgo~A) in earlier radio surveys, it is listed in the 3CRR catalogue as `A1552' (rather than having a `3C'/`4C' name). Our host-galaxy identification is in agreement with \citet{Owen1992}. 


{\bf G4Jy~1034} (GLEAM~J125437$-$123333; 3C~278) may not be as extended as other sources, but the high density of mid-infrared positions makes it difficult to identify the correct host-galaxy by eye. This WAT radio-galaxy is also known as NGC~4782, in a common envelope with NGC~4783, and the 5-GHz image presented by \citet{Baum1988} allows us to identify the radio core. As such, we update the AllWISE position to that of the 6dFGS source, g1254357$-$123407, in agreement with the literature \citep[e.g.][]{Morganti1993}.


The WAT radio-galaxy, {\bf G4Jy~1067} (GLEAM\,J132606$-$272641, GLEAM\,J132616$-$272632), is PKS~B1323$-$271 in cluster Abell~1736. Whilst the identification provided by \citet{vanVelzen2012} is the likely host galaxy, we cannot find a high-resolution radio image to confirm the core position. Therefore we leave the host flag as `u'.

{\bf G4Jy~1094} (GLEAM~J134855$-$252700) is a WAT radio-galaxy in cluster Abell~1791 (Figure~\ref{benttailoverlays}e). We use a 4.9-GHz image \citep{Gregorini1994} to confirm that the correct host galaxy is the AllWISE source that is also in 6dFGS (g1348542$-$252724).

TGSS contours show that {\bf G4Jy~1245} (GLEAM~J152007$-$283411; PKS 1517$-$283), in Abell~3618, is a WAT radio-galaxy. Based upon the 4.9-GHz contours presented by \citet{Gregorini1994}, we identify the 6dFGS source (g1520060$-$283420) as the host galaxy, lying at the mid-point between the two TGSS detections (Figure~\ref{benttailoverlays}f).  


{\bf G4Jy~1496} (GLEAM~J183626+193946, GLEAM~J183640+194318, GLEAM~J183649+194105) is the WAT radio-galaxy, PKS~B1834+19 (figure~6e of Paper~I). It has an unambiguous mid-infrared source coinciding with the peak emission in SUMSS (which marks the core), with our host-galaxy identification in agreement with \citet{vanVelzen2012}. 

The last bent-tail galaxies that we identify in the G4Jy Sample are {\bf G4Jy~1704} and {\bf G4Jy~1705}. These are discussed and shown in appendix~D.3 of Paper~I. 

\subsubsection{Head-tail radio-galaxies}
\label{subsec:headtails}

\begin{table}
\centering 
\caption{A list of G4Jy sources that have head-tail morphology, which we label as `complex' in the G4Jy catalogue (see Sections~\ref{sec:morphology} and \ref{subsec:headtails}). For these sources the host-galaxy identification is at the `head' of the radio emission.}
\begin{tabular}{@{}rcc@{}}
\hline
Source  & GLEAM component(s)   \\
\hline
G4Jy~100  & GLEAM~J005559$-$012139  \\
G4Jy~101  & GLEAM~J005623$-$011742  \\ 
G4Jy~110  & GLEAM~J010045$-$503129  \\
G4Jy~150 & GLEAM~J012548$-$012157  \\
G4Jy~204 & GLEAM~J015620+053724  \\  
 G4Jy~325  & GLEAM~J030653$-$120627  \\   
 G4Jy~466 & GLEAM~J043014$-$613244  \\ 
 G4Jy~475 & GLEAM~J043409$-$132250  \\ 
 & GLEAM~J043415$-$132717 \\  
G4Jy~595 & GLEAM~J060849$-$655110  \\ 
G4Jy~684 & GLEAM~J080535$-$005813  \\  
G4Jy~812 & GLEAM~J100140$-$305823  \\   
G4Jy~935 & GLEAM~J113943$-$464032  \\
 & GLEAM~J113956$-$463743  \\
G4Jy~984 & GLEAM~J121740+033940 \\   
G4Jy~1504 & GLEAM~J184315$-$483638  \\   
G4Jy~1606 & GLEAM~J201126$-$564322 \\
G4Jy~1638 & GLEAM~J203444$-$354849  \\   
G4Jy~1714 & GLEAM~J214014$-$441238  \\    
G4Jy~1817  & GLEAM~J231912$-$420614  \\
\hline
\label{listofheadtails}
\end{tabular}
\end{table}

\begin{figure*}%
\centering
\subfigure[G4Jy~100 and G4Jy~101]{
	\includegraphics[scale=1.1]{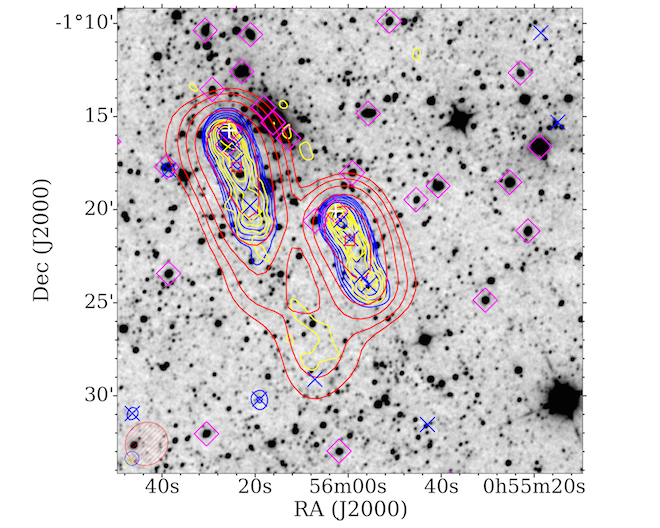} 
	} 
\subfigure[G4Jy~204]{
	\includegraphics[scale=1.1]{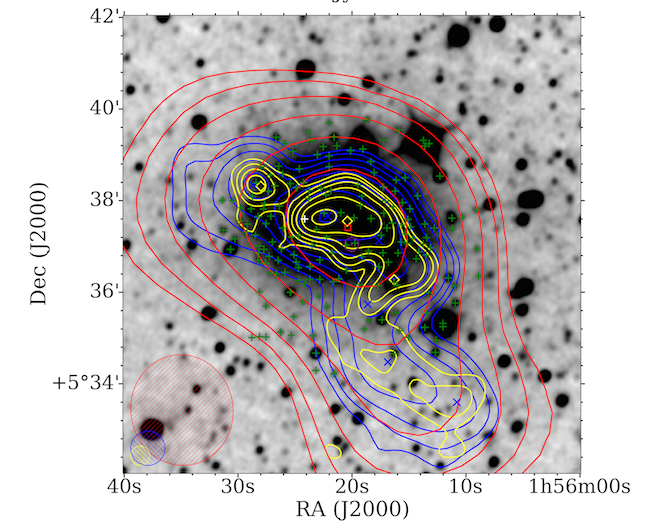} 
	} 
\subfigure[G4Jy~325]{
	\includegraphics[scale=1.1]{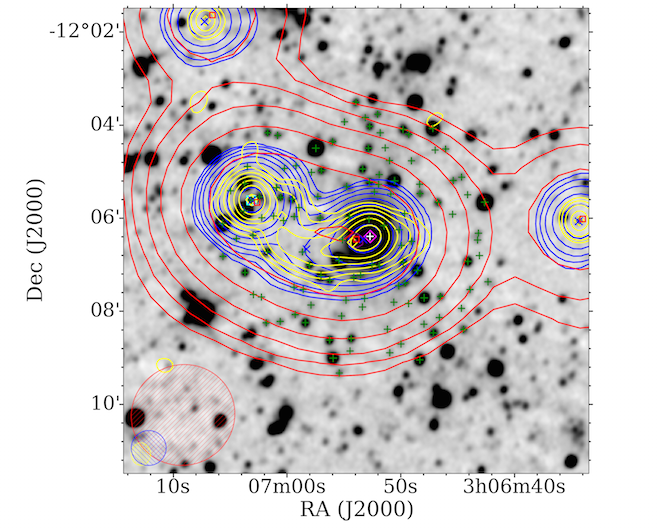} 
	} 
\subfigure[G4Jy~1638]{
	\includegraphics[scale=1.1]{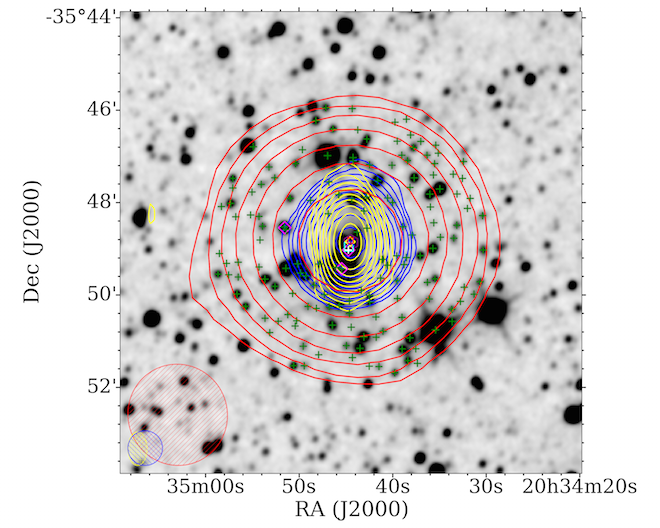} 
	} 
\caption{Examples of head-tail radio-galaxies in the G4Jy Sample (Section~\ref{subsec:headtails}). The datasets, contours, symbols, and beams are the same as those used for Figure~\ref{FlameNebulaImage}. Host galaxies are highlighted with a white plus, and for the 10$'$ overlays, all AllWISE positions within 3$'$ of the centroid are also shown (green plus signs). \label{headtailoverlays} }
\end{figure*}

In proximity (and parallel) to one another are the head-tail galaxies {\bf G4Jy~100} (GLEAM~J005559$-$012139) and {\bf G4Jy~101} (GLEAM~J005623$-$011742), shown in Figure~\ref{headtailoverlays}a. Their hosts are confirmed by FIRST images, which also begin to resolve the radio emission, showing NAT morphology.

{\bf G4Jy~150} (Section~\ref{sec:minkowski}; Figure~\ref{varietyoverlays}c) is the next head-tail galaxy in our sample, followed by {\bf G4Jy~204} (GLEAM~J015620+053724). The latter is NGC~742, in a common halo with NGC~741. We interpret the trail of low-frequency emission as associated with G4Jy~204, based on a 5-GHz image by \citet{Schellenberger2017}. This shows two radio jets emanating from the source and being strongly bent in a westward direction. In addition, the well-defined radio-core allows us to select the correct mid-infrared position in a dense field (Figure~\ref{headtailoverlays}b).

{\bf G4Jy~325} (GLEAM~J030653$-$120627) is within cluster Abell~415, and \citet{vanVelzen2012} identify this radio emission as PKS~B0304$-$12. However, PKS~B0304$-$12 is a blend of two radio sources: G4Jy~325 and an unrelated point-source (GLEAM~J030702$-$120539) towards the east and detected in AT20G (Figure~\ref{headtailoverlays}c). These are not distinguished in the catalogue of \citet{vanVelzen2012}, which explains why our total 1.4\,GHz flux-density is less than half the value they calculate (where their automated algorithm sums over three, rather than two, NVSS components). High-resolution radio images by \citet{ODea1985} and \citet{Owen1997} support our `head-tail' label, but despite our conflicting interpretation of the morphology, we agree with \citet{vanVelzen2012} as to the host-galaxy identification (g0306527-120624 at $z=0.079$). The difference is that we refer to G4Jy~325 alone \citep[B0304$-$123{\it A} in][]{ODea1985}, whilst they incorrectly associate G4Jy~325 with B0304$-$123{\it B} (whose host galaxy is AllWISE~J030703.13$-$120536.8).

\begin{figure*}
\centering
\subfigure[G4Jy overlay of G4Jy~466]{
	\includegraphics[scale=1.1]{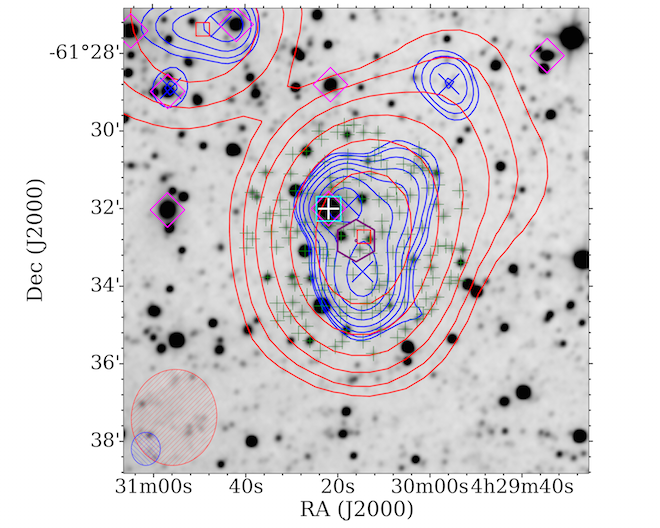}
	}
\subfigure[1.3\,GHz/optical overlay of G4Jy~466]{
	\includegraphics[scale=1.1]{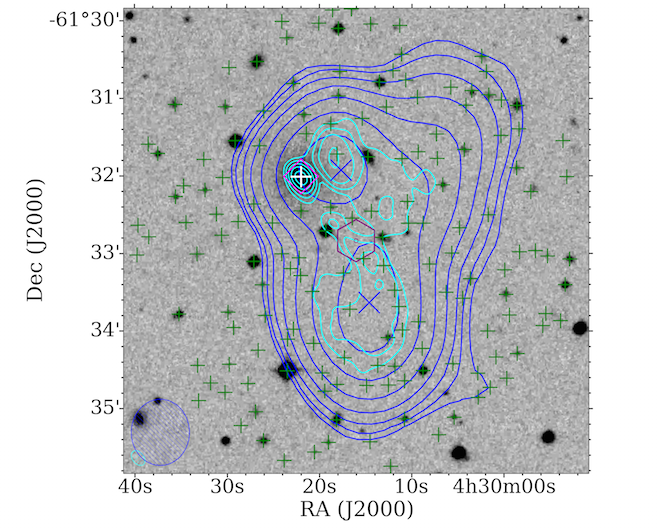}
	}
\caption{Two overlays for G4Jy~466, which is B0429$-$616 (Section~\ref{subsec:headtails}). The first overlay, (a), uses the same datasets, contours, symbols, and beams as those used for Figure~\ref{clusterrelic} (with the exception of TGSS, which is unavailable at this declination). The second overlay, (b), uses an optical image from SuperCOSMOS (\citealt{Hambly2001}; inverted greyscale) and a 1.3-GHz image (cyan contours) from \citet{ReidA1999} that was provided courtesy of Richard Hunstead. The latter was obtained using ATCA in its 6C configuration, resulting in a beam of $12.9'' \times 8.8''$ (cyan ellipse in the right-hand panel). Also overlaid are the same blue contours (using SUMSS) as in panel (a), for reference. For each set of contours in this figure, the lowest contour is at the 3\,$\sigma$ level (where $\sigma$ is the local rms), with the number of $\sigma$ doubling with each subsequent contour (i.e. 3, 6, 12\,$\sigma$, etc.). Positions from AllWISE are indicated by green plus signs, with the host galaxy highlighted in white. \label{ATCAheadtail}}
\end{figure*}

For {\bf G4Jy~466} (GLEAM~J043014$-$613244), also known as B0429$-$616 in cluster Abell~3266, we agree with \citet{Burgess2006b} as to the location of the radio core. This is at the position of the detections in AT20G and 6dFGS (g0430220-613201 at $z=0.055$), and confirmed by an ATCA image (in the thesis of \citealt{ReidA1999}), which we publish for the first time in Figure~\ref{ATCAheadtail}. At this higher resolution, G4Jy~466 appears as a WAT radio-galaxy, but we continue to refer to its morphology as `complex', in keeping with other (unresolved) head-tail galaxies in the sample.

{\bf G4Jy~475} (GLEAM~J043409$-$132250 and GLEAM~J043415$-$132717) is B0431$-$134 in cluster Abell~496.  A second `tail' is not clearly distinguished in our overlay (figure~6b of Paper~I), so we describe this as a head-tail galaxy and label its morphology `complex'. Our host-galaxy identification is in agreement with the core position, shown in a 4.9-GHz image by \citet{ODea1985}.  

\begin{figure*}
\centering
\subfigure[G4Jy overlay for G4Jy~595]{
        \includegraphics[scale=1.1]{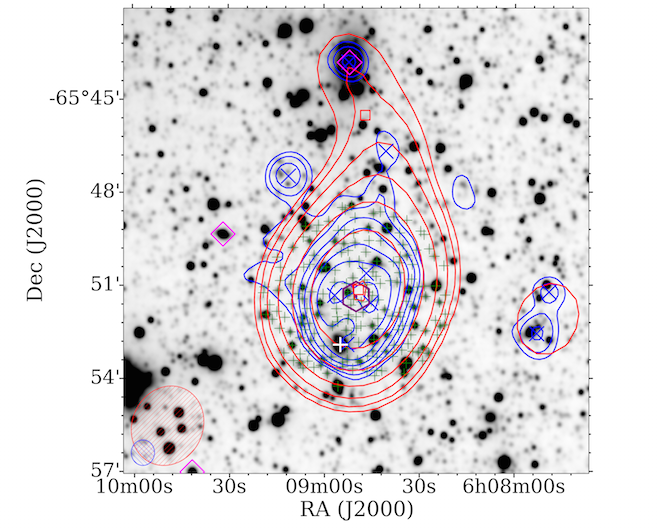}
	}
\subfigure[SUMSS image for G4Jy~595]{
	\includegraphics[scale=1.1]{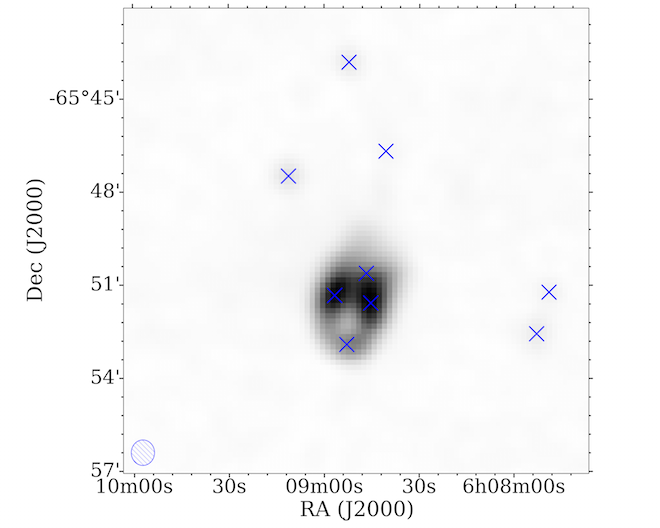}
	}
\caption{G4Jy~595 (Section~\ref{subsec:headtails}), as it appears (a) in the overlay used for visual inspection, and (b) in SUMSS (inverted greyscale). The overlay uses the same datasets, contours, symbols, and beams as those used for Figure~\ref{FlameNebulaImage}, but where blue contours and crosses represent SUMSS emission. Positions from AllWISE are indicated by green plus signs, with that corresponding to the host galaxy highlighted in white. \label{hollow}}
\end{figure*}

The SUMSS contours for {\bf G4Jy~595} (GLEAM~J060849$-$655110; B0608$-$658) show a strange morphology, with peaks in the radio emission along the east-west direction but extension in the north-south direction. It is only upon inspecting the SUMSS image alone (Figure~\ref{hollow}b) that we can appreciate that this is actually a NAT radio-galaxy, with a `hollow' in the radio emission near the core. Consequently, we update the mid-infrared host-galaxy in agreement with \citet{Jones1992}, but leave the morphology label as `complex' (since a `triple' morphology is not clear from the G4Jy overlay). In addition, the extended low-frequency emission towards the north suggests that GLEAM~J060847$-$654532 may be associated with G4Jy~595. However, it is unclear by how much the point source that is further north (coincident with a 6dFGS position) is potentially contributing towards this GLEAM component. Therefore, we continue with our conservative approach in listing G4Jy~595 as single-component in GLEAM.

{\bf G4Jy~935} (GLEAM~J113943$-$464032 and GLEAM~J113956$-$463743) is PKS~B1137$-$463 (figure~4d of Paper~I). Whilst the likely host-galaxy for this head-tail source is AllWISE~J113954.66$-$463749.3, we cannot find a high-resolution radio image to confirm this identification. Hence, we leave the host flag as `u'. 


{\bf G4Jy~1504} (GLEAM~J184315$-$483638; PKS~B1839$-$48) is shown by \citet{Morganti1993} to have WAT morphology. However, this is not clear in our overlay (with SUMSS artefacts causing distortion), so we label the morphology as `complex' (i.e. showing head-tail morphology). \citet{Morganti1993} also quote the core position, which agrees with our AllWISE identification (coincident with both the AT20G and 6dFGS positions).

 The BCG of Abell~3667 is another head-tail galaxy \citep{Goss1982}, appearing in the G4Jy Sample as {\bf G4Jy~1606} (GLEAM~J201126$-$564322; Figure~\ref{clusterrelic}). ATCA observations of this source (B2007$-$568) at 1.4\,GHz \citep{Riseley2015} corroborate our identification of the host galaxy. 

{\bf G4Jy~1638} (GLEAM~J203444$-$354849) is B2031$-$359, and appears as a head-tail galaxy in our overlay (Figure~\ref{headtailoverlays}d), hence our morphology label of `complex'. However, this source is shown to have NAT/WAT morphology in a 4.9-GHz image presented by \citet{Ekers1989}, which we use to confirm that g2034447$-$354902, in 6dFGS, is the host galaxy.

Finally, it is unclear whether {\bf G4Jy~1410} (GLEAM~J172437$-$024246; see appendix~D.3 of Paper~I) is a `double' or a head-tail galaxy (given the relative compactness of the TGSS contours, compared to what we see for other head-tails). Whilst the presence of a bright mid-infrared source (AllWISE~J172437.79$-$024305.6) at one end of the radio emission appears to support the `head-tail' interpretation, it could also be obscuring a host galaxy that is coincident with the TGSS/NVSS detection (which would lend credence to this source being a `double'). Therefore, we err on the side of caution by setting the morphology label to `complex' and leaving the host-galaxy flag as `u'.

\subsection{Known, giant radio-galaxies}
\label{sec:GRGs}

We follow the definition of a `giant radio-galaxy' \citep[GRG; e.g.][]{Willis1978} as being a radio galaxy that has a projected, linear size of $\geq1$\,Mpc. A summary of the properties of the GRGs described in this subsection can be found in Table~\ref{listofGRGs}. We emphasise that these are radio sources that are already known to be GRGs, as an extensive search for {\it new} GRGs is beyond the scope of this paper. 


\begin{table*}
\centering 
\caption{A list of G4Jy sources that are already known to be GRGs (Section~\ref{sec:GRGs}). Linear sizes are calculated using the angular sizes and redshifts below in conjunction with the online cosmology calculator, CosmoCalc \citep{Wright2006}$^1$.}
\begin{center}
\begin{threeparttable}
\begin{tabular}{@{}rlccccc@{}}
 \hline
Source & Other name    & Angular size & Reference for & Redshift & Reference for &  Linear size   \\
 &     &  / arcmin & the angular size &  & the redshift &   / Mpc  \\
 \hline
G4Jy~133 & B0114$-$47 &  9.5 & (1) & 0.146 & (2) & 1.5 \\  
G4Jy~347 & B0319$-$453 &  25.6 & (1) & 0.063 & (3)& 1.9 \\  
G4Jy~517 & B0503$-$286 &  35.7 & (4) & 0.038 & (2)  & 1.6 \\ 
G4Jy~1079 & B1331$-$099 &  13.5 & (4) & 0.084 & (2) & 1.3 \\ 
G4Jy~1279 & B1545$-$321 &  9.0 & (5) & 0.108 & (6) & 1.1 \\  
G4Jy~1282 & 3C 326 &  19.5 & (7) & 0.090 & (8) & 2.0\\   
G4Jy~1525 & B1910$-$800 &  6.1 & (9) & 0.346 & (9) & 1.8 \\   
G4Jy~1613 & B2014$-$558 &  20.0 & (1) & 0.061 & (2) & 1.4\\  
\hline
\label{listofGRGs}
\end{tabular}
\vspace{-3mm}
\begin{tablenotes}{\footnotesize 
\item{References: (1) \citet{Jones1992}, (2) \citet{Jones2009}, (3) \citet{Jones1989}, (4) Paper~I \citep{White2020a} alongside this work, (5) \citet{Saripalli2003}, (6) \citet{Simpson1993}, (7) \citet{Willis1978}, (8) \citet{Smith1976}, and (9) \citet{Subrahmanyan1996}. $^{1}$http://www.astro.ucla.edu/\textasciitilde wright/CosmoCalc.html}}
\end{tablenotes}
\end{threeparttable}
\end{center}
\label{listofGRGs}
\end{table*}

\begin{figure*}%
\centering
\subfigure[G4Jy~133]{
	\includegraphics[scale=1.1]{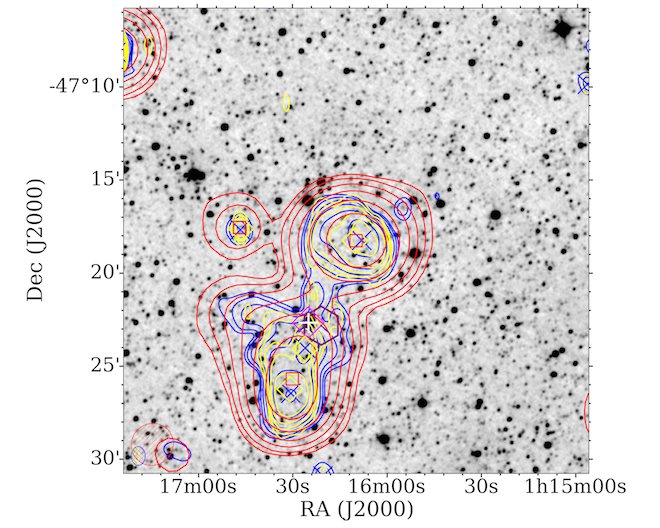} 
	} 
\subfigure[G4Jy~347]{
	\includegraphics[scale=1.1]{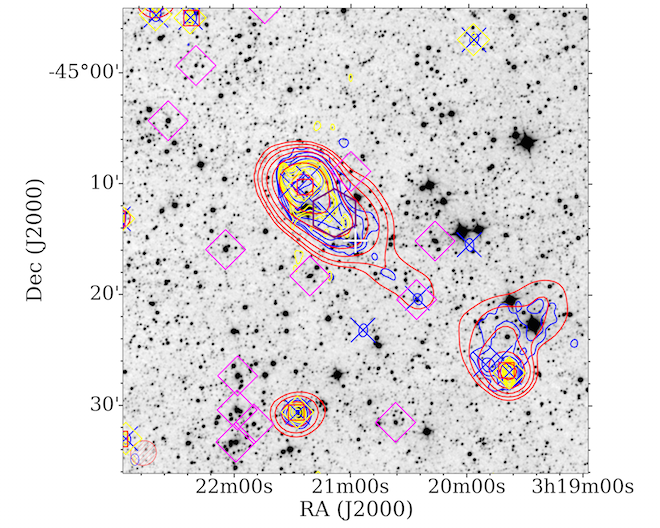} 
	} 
\subfigure[G4Jy~517]{
	\includegraphics[scale=1.1]{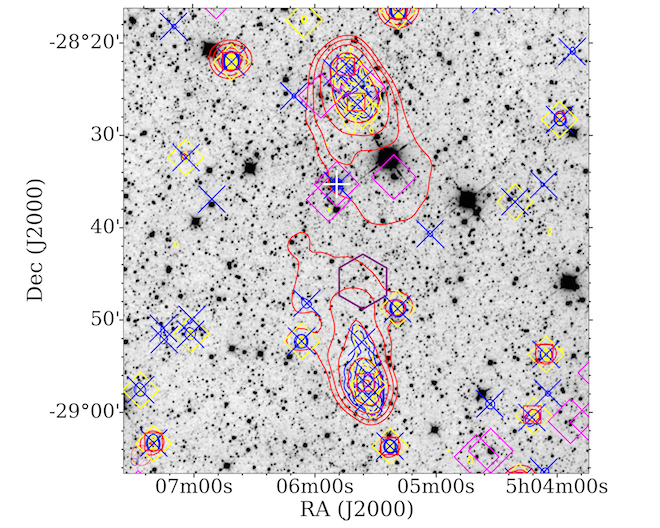} 
	} 
\subfigure[G4Jy~1079]{
	\includegraphics[scale=1.1]{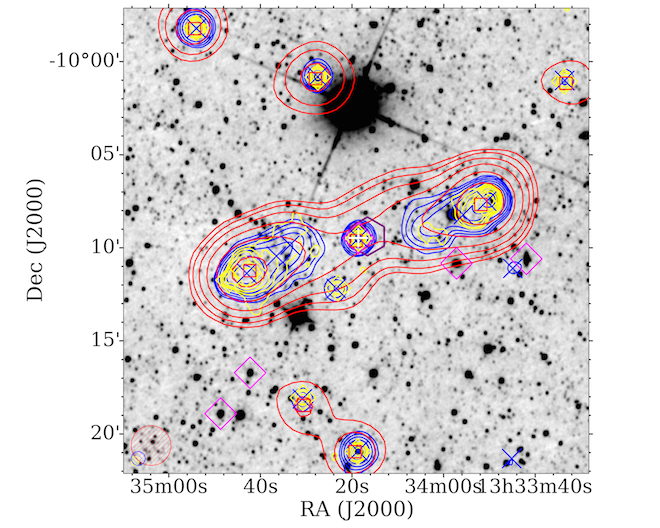} 
	} 
\subfigure[G4Jy~1279]{
	\includegraphics[scale=1.1]{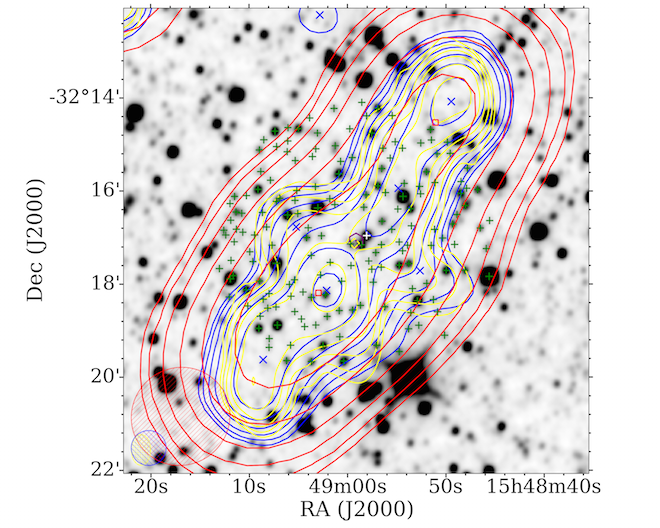} 
	} 
\caption{Five of eight known GRGs in the G4Jy Sample (Section~\ref{sec:GRGs}). The datasets, contours, symbols, and beams are the same as those used for Figure~\ref{FlameNebulaImage}, but where blue contours, crosses, and ellipses correspond to NVSS {\it or} SUMSS. Host galaxies are highlighted with a white plus, and for the 10$'$ overlay, all AllWISE positions within 3$'$ of the centroid are also shown (green plus signs). \label{GRGoverlays} }
\end{figure*}

The SUMSS contours for {\bf G4Jy~133} (GLEAM~J011609$-$471816, GLEAM~J011630$-$472542; Figure~\ref{GRGoverlays}a) suggest the presence of inner jets, which simplifies the identification of the host galaxy. Although the 6dFGS detection (g0116251$-$472241) appears to be the obvious candidate, we note the density of mid-infrared sources nearby. We therefore check against the literature \citep{Danziger1983,Jones1992}, where this source is also referred to as PKS~B0114$-$47. This confirms that the 6dFGS position is correct, and that the radio source is at $z=0.146$. Its angular extent of 9$\farcm$5 therefore corresponds to a physical size of 1.5\,Mpc, hence its qualification as a GRG. 

From Figure~\ref{GRGoverlays}b it may be thought that the 6dFGS detection (g0320261$-$452029), near the mid-point between the two lobes, marks the host galaxy for {\bf G4Jy~347} (GLEAM~J031939$-$452649 and GLEAM~J032123$-$451021). However, we note from the literature \citep[B0319$-$453 in][]{Jones1992,Saripalli1994} that the host galaxy is actually ESO 248-G10, {\it 3.1\,arcmin from the centroid position}. This is beyond our search radius in the AllWISE catalogue, and so (for re-inspection) we also include mid-infrared sources within 3$\farcm$0 of the host-galaxy position. A spectral-index map by \citet{Safouris2009} shows flat-spectrum radio emission associated with the core, and so provides further confirmation that the host galaxy has been identified correctly. Being at $z=0.063$ \citep{Jones1989}, the 25$\farcm$6 separation of the radio lobes corresponds to a projected size of 1.9\,Mpc, making this source another GRG. 

The lobes of {\bf G4Jy~517} (GLEAM~J050535$-$285648, GLEAM~J050539$-$282627, GLEAM~J050544$-$282236) show distinct polarised morphology between 216 and 1400\,MHz (\citealt{Riseley2018}; Riseley et al., submitted), and are $\sim36'$ apart. Such a large spatial separation means that identifying the host galaxy can become more difficult, owing to the substantial number of potential candidates (Figure~\ref{GRGoverlays}c). To help resolve this we again consult the literature, where `B0503$-$286' has previously been used to refer to both the entire radio-galaxy and, individually, its southern lobe. \citet{Saripalli1986} reveal that the host galaxy (ESO 422-G028; g0505492$-$283519 in 6dFGS) is a $B = 15$\,mag elliptical at $z=0.038$. This means that this radio source is another GRG, having a physical size of 1.6\,Mpc. As shown in Figure~\ref{GRGoverlays}c, the combination of large angular-size and asymmetric radio-lobes results in a host galaxy that is {\it 10.9\,arcmin from the centroid position}. Therefore, like for G4Jy~347, we supplement our subset of the AllWISE catalogue by adding mid-infrared sources in the vicinity of the optical position (again using a search radius of 3$\farcm$0 for consistency). After adding these positions to the `close-up' overlay (now extended to 20$'$ across; see section~5.1 of Paper~I), this source is ready for re-inspection regarding mid-infrared identification. Follow-up radio observations at higher frequencies \citep{Subrahmanyan2008} confirm the position of the core, and so we are satisfied that identification of the host galaxy is correct. 

{\bf G4Jy~1079} is an archetypal `triple', with three distinct components in GLEAM (GLEAM~J133351$-$100740, GLEAM~J133419$-$100937 and GLEAM~J133442$-$101114). It has an obvious core \citep{Saripalli1996}, but this still coincides with two positions in the AllWISE catalogue (Figure~\ref{GRGoverlays}d). However, we find no reference to two galaxies in the literature (nor evidence in SuperCOSMOS), but \citet{Schilizzi1975b} notes that the elliptical host resides in `a very red irregularly shaped envelope'. Therefore, the two AllWISE positions may be a consequence of fitting the host galaxy's extended structure. Proceeding with the corresponding identification in 6dFGS (g1334186$-$100929, with $z=0.084$), we determine a physical size of 1.3\,Mpc for G4Jy~1079.

{\bf G4Jy~1279} (GLEAM~J154851$-$321431 and GLEAM~J154902$-$321811) is the re-started GRG also known as B1545$-$321 (Figure~\ref{GRGoverlays}e). Evidence for renewed AGN activity is provided by high-resolution ATCA images \citep{Saripalli2003}, which show a pair of inner jets embedded within the older, outer radio-lobes. They also show that neither of the two AllWISE sources closest to the centroid are correct, and instead confirm the existing optical identification that we find in the literature \citep{Jones1992,Simpson1993}. We therefore manually update our choice of the host-galaxy position accordingly. 

Having cross-identified {\bf G4Jy~1282} (GLEAM~J155120+200312, GLEAM~J155147+200424, and GLEAM~J155226+200556) as 3C~326 (section~5.4 of Paper~I), we refer to the literature for the host-galaxy identification. Multi-frequency observations by \citet{Willis1978} show evidence of a weak core between the two radio lobes, coincident with the southern member\footnote{Note that this is in agreement with \citet{Laing1983} and the online 3CRR catalogue (https://3crr.extragalactic.info), but differs from the 3CRR catalogue that is available through VizieR (http://vizier.u-strasbg.fr). The latter currently quotes an optical position for the northern member of the galaxy pair. Similarly, for 4C~+13.66 (G4Jy 1456) and 3C 437 (G4Jy 1724), we favour the up-to-date positions from the online version of the 3CRR catalogue over those provided through VizieR. These host-galaxy identifications are supported by \citet{Rawlings1996} and \citet{Best1997}, respectively.} of a pair of galaxies at $z = 0.0895$ \citep{Smith1976}. With lobes separated by 19$\farcm$5, this means that G4Jy~1282 is 2\,Mpc in size.

{\bf G4Jy~1525} (GLEAM~J191905$-$795737 and GLEAM~J191931$-$800128) is B1910$-$800. We agree with \citet{Subrahmanyan1996} that the optical identification of \citet{Jones1992} is incorrect, and that it should be the optically-faint galaxy that is on-axis at the `neck' of the SUMSS contours (connecting the two lobes). This host appears quite bright in the AllWISE image (figure~4e of Paper~I), and our host-galaxy position agrees with the core position \citep{Saripalli2005}. The source is at $z=0.346$ \citep{Subrahmanyan1996}, meaning that this extended `double' is 1.8\,Mpc across.  

The final GRG listed in Table~\ref{listofGRGs} is {\bf G4Jy~1613}, with distinct X-shaped morphology (Figure~\ref{Xshape}e), and so previously described in Section~\ref{sec:Xshapedsources}.

\subsection{Unclassified, extended radio-sources}
\label{sec:unclassifiable}

In this section we describe three sources that are difficult to classify, on account of their unusual morphology. To investigate further, they (along with 137 other G4Jy sources) will be followed up with MeerKAT (PI: White). This instrument has excellent sensitivity to diffuse emission at 1.3\,GHz and will provide $\sim5''$ resolution images.


\subsubsection{A radio lobe, or cluster-related emission?}

We note extended low-frequency emission near to GLEAM~J140421$-$340018 (Figure~\ref{unclassified}a) that suggests that this is another case of identifying just one of the lobes of a powerful radio-galaxy (cf. appendix~D.1 of Paper~I). Whilst exploration of the surrounding region (using the full GLEAM mosaic, rather than a 1\arcdeg\ cutout) shows no sign of a second radio-lobe, there are hints of such a counterpart in a deep MWA observation. Further observations are needed to confirm this. Meanwhile, SUMSS emission for this GLEAM component appears in the catalogue of \citet{vanVelzen2012}, where it has been associated with NGC~5419 (GLEAM~J140338-335848, $S_{\mathrm{151\,MHz}}=1.4$\,Jy). However, the {\it direction} of the suggested lobe (with respect to NGC~5419) then appears to be incongruous with the extended emission in GLEAM. 

An alternative explanation is that GLEAM~J140421$-$340018 is dominated by emission from a cluster relic, with nearby GLEAM components (GLEAM~J140323$-$341020, GLEAM~J140339$-$340430,  GLEAM~J140356$-$341616, GLEAM~J140403$-$341304, GLEAM~J140420$-$340904) being associated emission from a cluster radio-halo. See \citet{Subrahmanyan2003} for further discussion of this source (B1400$-$33) and an image of the radio emission at 330\,MHz using the VLA.

Due to the difficulty in classifying GLEAM~J140421$-$340018 (Johnston--Hollitt et al., in preparation), we label it as `complex' (Section~\ref{sec:morphology}), treat it as single-component in GLEAM (going by the source-name {\bf G4Jy~1117}), and do not provide a mid-infrared identification.

\subsubsection{A possible GRG with `triple' morphology}
\vspace{1mm} 

\begin{figure}
\centering
\subfigure[G4Jy~1117]{
	\includegraphics[scale=1.1]{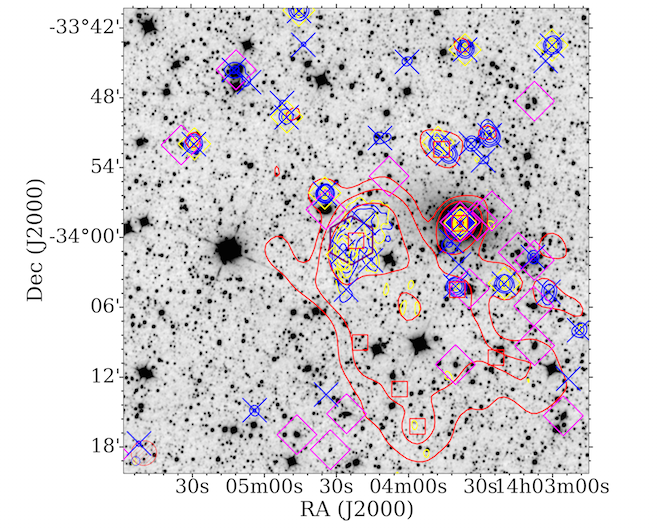}
	}
\subfigure[G4Jy~113]{
	\includegraphics[scale=1.1]{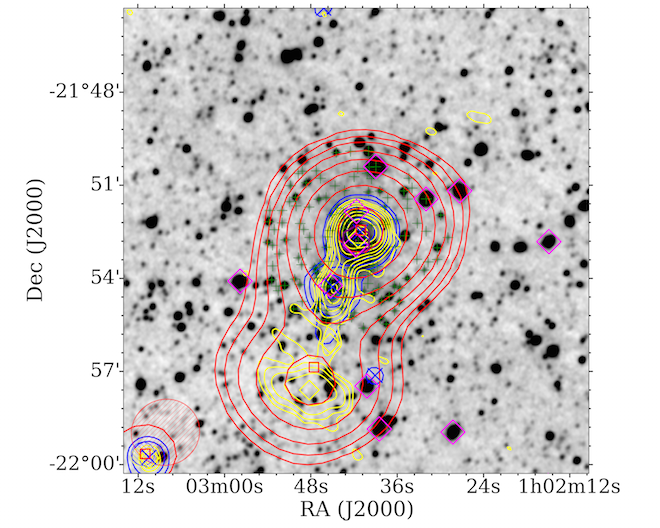}
	} 
\subfigure[G4Jy~513]{
	\includegraphics[scale=1.1]{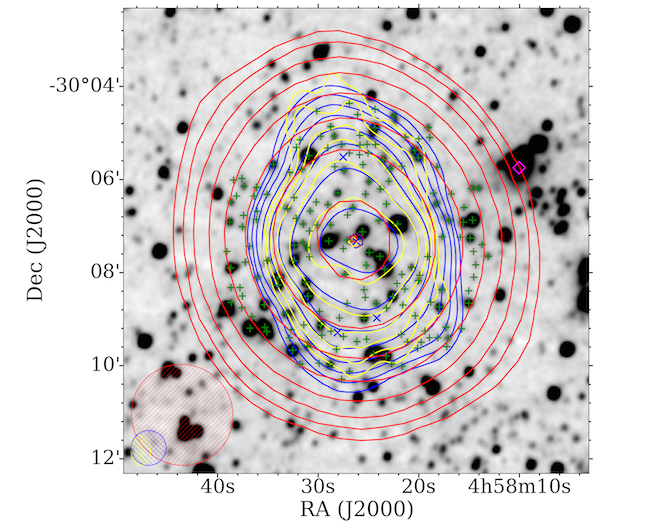}
	} 
\caption{Overlays for three G4Jy sources whose morphology we are unable to classify (Section~\ref{sec:unclassifiable}). The datasets, contours, symbols, and beams are the same as those used for Figure~\ref{FlameNebulaImage}. \label{unclassified}}
\end{figure}

We treat {\bf G4Jy~113} (GLEAM~J010241$-$215227) in the same way as G4Jy~1117, since understanding its morphology is also not straightforward (Figure~\ref{unclassified}b). We find that it is located in cluster Abell~133, with \citet{Slee2001} interpreting the steep-spectrum radio emission as a cluster relic -- like G4Jy~1605 (Section~\ref{sec:clusterrelics}) -- generated via merger shocks. They dismiss the suggestion by \citet{Rizza2000} that it is a remnant radio-lobe, although this second interpretation is supported by \citet{Fujita2002}, who use a combination of radio and X-ray observations to infer the energetics of the system.

Pursuing the idea that the radio emission originated from a radio galaxy, the next question concerns the location of the host. At the centre of Abell~133 is ESO 541-G013 (g0102418$-$215256 at $z = 0.057$), which coincides with compact radio-emission. \citet{Rizza2000} see a bridge of emission that connects it with the extended emission (i.e. `lobe') in the north, implying that ESO 541-G013 may be the former host-galaxy. Meanwhile, \citet{Slee2001} do not see this bridge in their observations.

However, the radio images presented by \citet{Rizza2000} and \citet{Slee2001} are at 1.4\,GHz. The picture becomes more unusual when we consider the low-frequency morphology, as shown by the TGSS contours (Figure~\ref{unclassified}b). If `triple' is now the correct interpretation of this radio emission, then the radio galaxy is much more extended than previously thought, spanning 6$'$. In this scenario, GLEAM~J010247$-$215651 (the southern lobe) is associated with GLEAM~J010241$-$215227, and the host is likely the compact radio-source, g0102453$-$215414 at $z = 0.293$ [referred to as galaxy `J' by \citet{Slee2001} in their figure~6]. This redshift corresponds to a linear scale of 263\,kpc/arcmin (which would make G4Jy~113 a GRG), whilst the linear scale for the cluster in the foreground is 66\,kpc/arcmin. The next question is why the southern lobe has a steeper spectral-index than the northern lobe (as implied by the lack of NVSS emission for the southern lobe). Alternatively, each of the `segments' of low-frequency emission could be connected to unrelated sources.

\subsubsection{Cluster halo, or a remnant radio-galaxy?}

The amorphous appearance of {\bf G4Jy~513} (GLEAM~J045826$-$300717; PKS~B0456$-$30), in cluster Abell~3297, leads us to label the morphology as `complex' and the host flag as `u' (Figure~\ref{unclassified}c). \citet{Jones1992} suggest that the extended emission could be a cluster halo, but they also report an identification (which we are unable to verify). Meanwhile, \citet{Yuan2012} estimate an upper limit for the core flux-density ($<6.7$\,mJy at 5\,GHz), so it is possible that the radio emission is associated with a radio galaxy that has become inactive. Also, we note the uniformity of the TGSS and NVSS contours, and calculate that the spectral index between 151 and 1400\,MHz (`G4Jy\_NVSS\_alpha'; see section~6.6 of Paper~I) is $-0.68\pm0.01$.


\subsection{Sources with faint mid-infrared hosts}
\label{sec:fainthosts}

\begin{figure*}
\centering
\subfigure[G4Jy~180]{
	\includegraphics[scale=1.1]{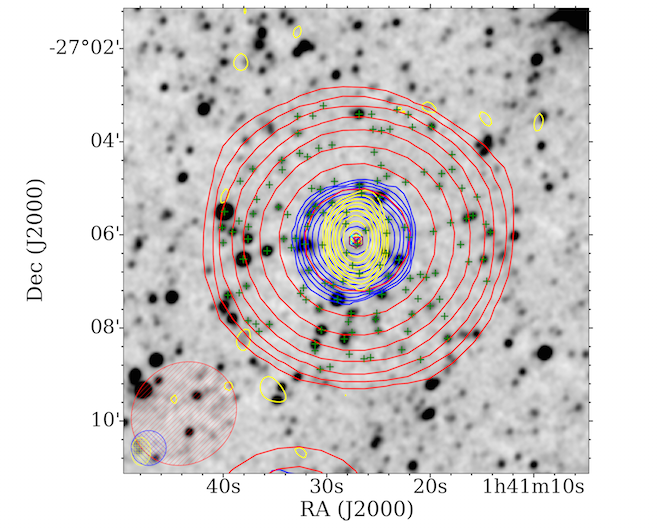}
	}
\subfigure[G4Jy~877]{
	\includegraphics[scale=1.1]{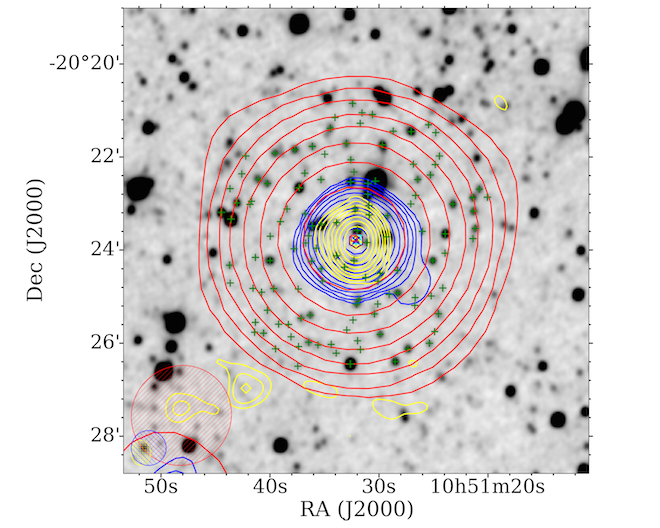}
	} 
\caption{Examples of G4Jy sources that have (relatively) faint mid-infrared hosts (Section~\ref{sec:fainthosts}). The datasets, contours, symbols, and beams are the same as those used for Figure~\ref{FlameNebulaImage}, with AllWISE positions within 3$'$ of the centroid (purple hexagon) indicated by green plus signs. \label{hostflagm}}
\end{figure*}

126 G4Jy sources have been assigned `m' for their host flag (Section~\ref{sec:identifyhost}), indicating that their radio position looks secure but no appropriate identification can be made using AllWISE. For example, 14 of these are sources where an AT20G detection indicates the location of the radio core, but the host galaxy is too faint in the mid-infrared to be characterised for the AllWISE catalogue (Figure~\ref{hostflagm}). These radio sources are likely to be at high redshift, and we confirm that this is the case for, at least, {\bf G4Jy~180} (GLEAM~J014127$-$270606; PKS~B0139$-$273) at $z = 1.440$ and {\bf G4Jy~877} (GLEAM~J105132$-$202344; PKS~B1049$-$201) at $z = 1.116$ \citep{McCarthy1996,Mahony2011}.

As shown by \citet{Mahony2011}, the fraction of radio sources with an optical identification falls as the spectral index (which they measure between 5 and 20\,GHz) steepens. This is thought to indicate the presence of a high-redshift population of ultra-steep-spectrum sources ($\alpha < -1.3$), with numerous candidates \citep[e.g.][]{deBreuck2000} being followed up spectroscopically to confirm this. Indeed, this is the method exploited by \citet{Saxena2018}, who use the combination of TGSS (150\,MHz) and the VLA (1.4\,GHz) to identify the highest-redshift powerful radio-galaxy known, at $z = 5.72$.

In the search for high-redshift galaxies, an alternative to the method above (which relies on the radio spectral-index) is to identify infrared-faint radio sources (IFRSs). These were first classified by \citet{Norris2006}, with formal criteria for their selection put forward by \citet{Zinn2011}. The first of these criteria isolates sources with a high ratio of radio to mid-infrared flux-density, whilst the second criterion aims to remove low-redshift, radio-loud AGN via a mid-infrared flux-density cut. Most recently,  \citet{Orenstein2019} have applied this method to the Unified Radio Catalogue \citep[URC;][]{Kimball2008,Kimball2014}, in combination with data from AllWISE \citep{Cutri2013} and SDSS DR12 \citep{Alam2015}. This allowed them to spectroscopically confirm 108 IFRSs, with the highest-redshift IFRS being found at $z = 4.387$.

Regarding the G4Jy Sample, we will provide existing redshift information, and analysis of multi-wavelength data, in the next paper (Paper~III; White et al., in preparation). This will later be supplemented by new measurements from the Taipan Galaxy Survey \citep{daCunha2017}, which will provide optical spectroscopy over the entire southern-hemisphere. In addition, we will use the Southern African Large Telescope to observe the optically-fainter sources (PI: White).

\section{Additional literature checks}
\label{sec:additionalchecks}

In the previous section we highlighted some of the more-unusual radio sources in the G4Jy Sample, some of which fall into multiple categories of interest. In this section we record similar cross-checking against the literature, but for `normal' radio-galaxies, which compose the bulk of the sample. These checks have been the most time-consuming and labour-intensive step in our host-galaxy identification, but will hopefully prove to make the G4Jy catalogue (section~6 of Paper~I) a valuable resource for: (i) multi-wavelength studies of AGN across the southern sky, and (ii) assessing the performance of automated/machine-learning algorithms for cross-identification.

\subsection{Radio galaxies requiring further work}
\label{sec:ambiguous}


{\bf G4Jy~14} (GLEAM~J000707+053607) has two mid-infrared sources at a similar distance from the centroid position, both of which are plausible candidates for hosting the `double' radio emission. The radio lobes appear in FIRST but there is no detection of the core. Therefore, we are unable to resolve the ambiguity as to the host, and so set the host flag to `u' (Section~\ref{sec:identifyhost}).

We use FIRST to confirm that {\bf G4Jy~22} (GLEAM~J001310+005139) is a `double', and that we have correctly identified the host galaxy. However, the morphology is unresolved in our overlays, so we label the source as a `single' for consistency.

`Double' morphology is evident in the TGSS contours for {\bf G4Jy~23} (GLEAM~J001356$-$091952), which is also known as PKS~B0011$-$096. \citet{ReidR1999} present a 4.7-GHz image that distinguishes the lobes, but it is a FIRST image that confirms that the bright mid-infrared source nearest to the GLEAM position is the host.

\begin{figure}
\centering
\includegraphics[scale=1.1]{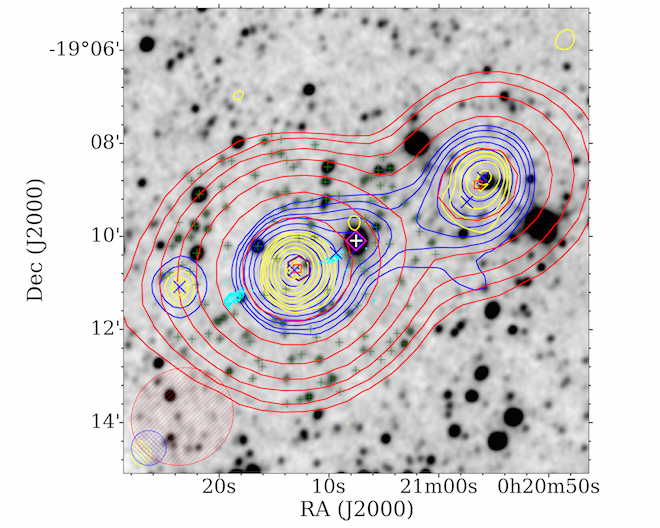}
\caption{An overlay for G4Jy~40 (Section~\ref{sec:ambiguous}). The datasets, contours, symbols, and beams are the same as those used for Figure~\ref{FlameNebulaImage}. In addition, positions from AllWISE are indicated by green plus signs, with that corresponding to the host galaxy highlighted in white. The cyan contours are from a 4.7\,GHz image \citep{ReidR1999} that is believed to have incorrect co-ordinates$^{\ref{rkp1999note}}$. }
\label{wrongcoordinates}
\end{figure}

Our radio contours for GLEAM~J002056$-$190853 and GLEAM~J002112$-$191041, in combination with the optically-bright source (g0021075$-$191006 in 6dFGS) lying between them, give the impression that these two GLEAM components are the lobes of an extended radio-galaxy. This is the interpretation of, for example, \citet{Nilsson1998}, who base their largest-angular-size measurement (of 252$''$) on the radio map of \citet{Schilizzi1975a}. However, this source ({\bf G4Jy~40}; PKS~B0018$-$19) is shown at 4.7\,GHz \citep{ReidR1999} to have `triple' morphology that spans $\sim160''$ from lobe to lobe (Figure~\ref{wrongcoordinates}). Our investigation is complicated by some of the images from \citet{ReidR1999} [the one for PKS~B0018$-$19 included] appearing to have incorrect co-ordinates, offset in different directions with respect to the J2000 reference-frame\footnote{\label{rkp1999note}Other sources, whose 4.7-GHz images are similarly affected, are PKS~B0350$-$07 (G4Jy~392), PKS~B1434$-$076 (G4Jy~1180) and PKS~B1453$-$10 (G4Jy~1209). Within this small subset, we see no trend between the direction of the offset and whether the epoch is quoted as `2000' or `1950' in the header of the image. We take the apparent offsets into consideration when identifying the host galaxy of the radio emission. The 4.7-GHz images for PKS~B0338$-$343 (G4Jy~376) and PKS~B1011$-$282 (G4Jy~829), however, show morphology that is consistent with the NVSS and TGSS images (J2000).}. We assume that the south-eastern hotspot at 4.7\,GHz should actually coincide with the NVSS component 002113$-$191043, which is supported by the location of polarised emission shown for `01-04' in appendix~A of \citet{Riseley2018}. On this basis, the core then aligns with the 6dFGS source, and so we provide AllWISE~J002107.53$-$191005.4 as the mid-infrared identification in AllWISE. This is in agreement with the host-galaxy identification for PKS~B0018$-$19 in the literature (see SIMBAD\footnote{http://simbad.u-strasbg.fr/simbad/}). In addition, until a high-resolution radio-image {\it with a large field-of-view} becomes available, we regard GLEAM~J002056$-$190853 as unassociated with G4Jy~40.

{\bf G4Jy~61} (GLEAM~J003354$-$073019) is PKS~B0031$-$07, and we update our choice of AllWISE host in agreement with \citet{Sadler2019}. Although further from the centroid (which may be influenced by Doppler boosting), this host galaxy is closer to the mid-point between the two NVSS components.

A `triple' morphology in FIRST confirms that, for {\bf G4Jy~62} (GLEAM~J003419+011851), the mid-infrared source closest to the centroid is the host galaxy. However, the FIRST image also reveals that the northern-most TGSS position marks the host of an unrelated `double'. As such, we update the confusion flag for G4Jy~62 to `1'.

The spatial resolution of the radio images makes it difficult to determine whether {\bf G4Jy~68} (GLEAM~J003744+131953) should be labelled `single' or `double'. As a result, there is ambiguity as to which mid-infrared source is the correct host-galaxy. However, this source appears in the 3CRR catalogue as 3C~16 (a `double'), and was observed at 0.75$''$ resolution by \citet{Gilbert2004}. They show the position of the radio core, allowing us to identify the appropriate mid-infrared host. 

The radio core is visible in the FIRST image for {\bf G4Jy~72} (GLEAM~J003918+031947), and confirms that the AllWISE position closest to the centroid is the correct selection. 

The 6dFGS source (g0046050$-$633319) appears to be the host galaxy of {\bf G4Jy~84} (GLEAM~J004601$-$633348) but there is another AllWISE candidate, which is on the radio axis of the `double' and a similar distance from the centroid. As we cannot find a high-resolution image in the literature to confirm the host, we leave the host flag as `u'.

{\bf G4Jy~121} (GLEAM~J010556+062814) is 4C~+06.06, and we update the AllWISE position to that lying on-axis between the two TGSS and NVSS detections.

\begin{figure*}
\centering
\subfigure[G4Jy~224]{
	\includegraphics[scale=1.1]{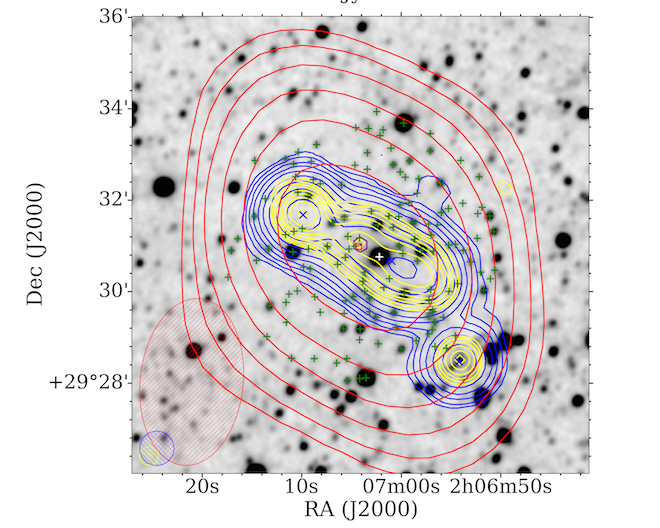}
	}
\subfigure[G4Jy~392]{
	\includegraphics[scale=1.1]{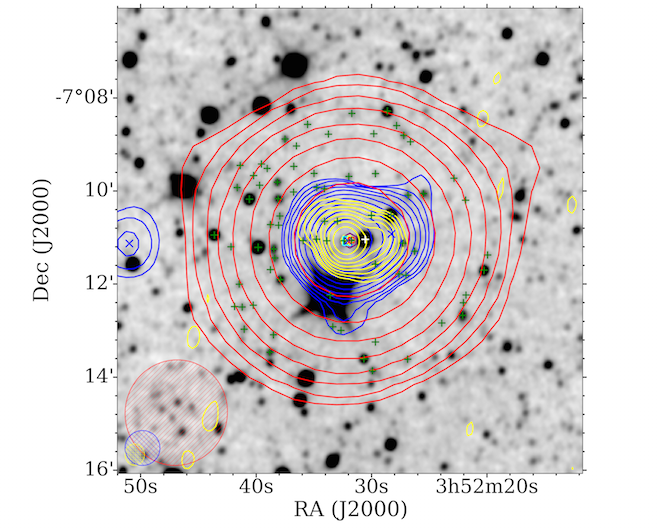}
	} 
\subfigure[G4Jy~499]{
	\includegraphics[scale=1.1]{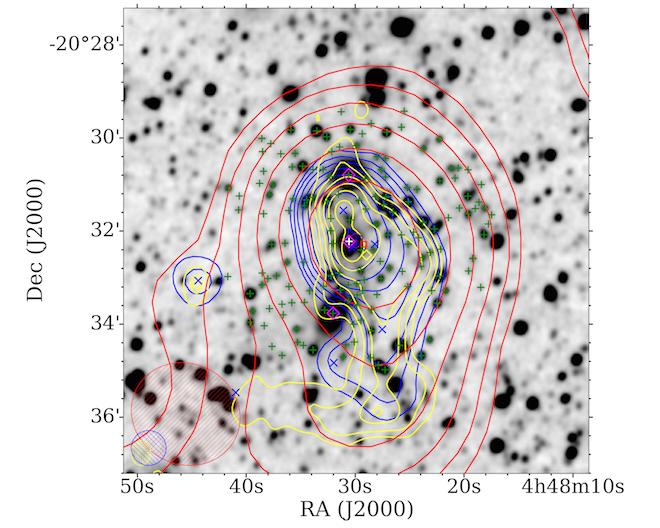}
	} 
\subfigure[G4Jy~680]{
	\includegraphics[scale=1.1]{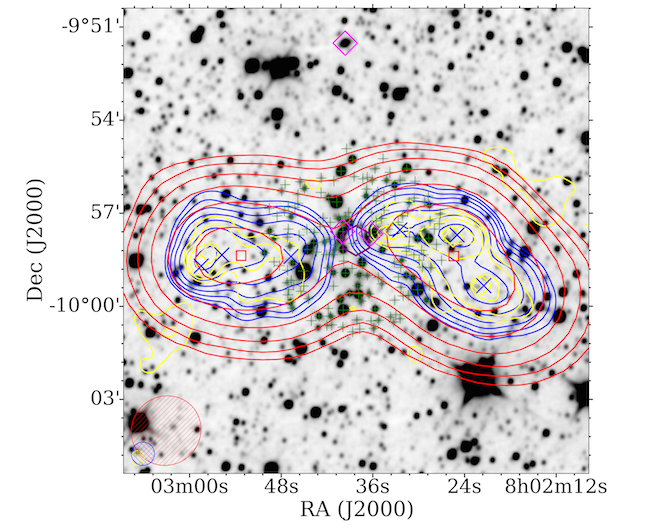}
	} 
\subfigure[G4Jy~886]{
	\includegraphics[scale=1.1]{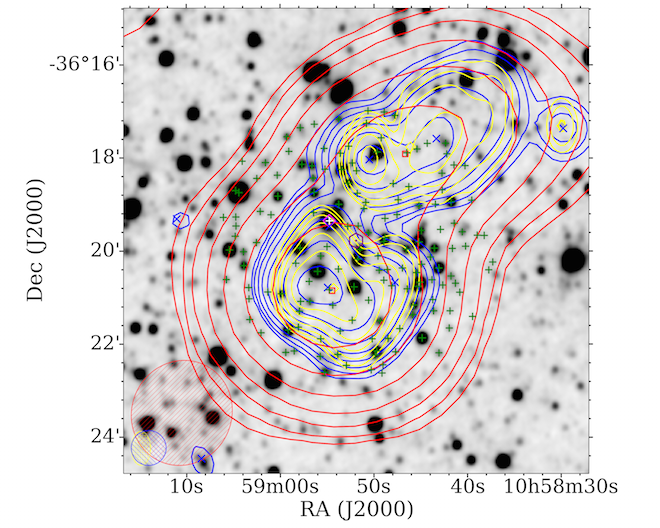}
	}
\subfigure[G4Jy~1003]{
	\includegraphics[scale=1.1]{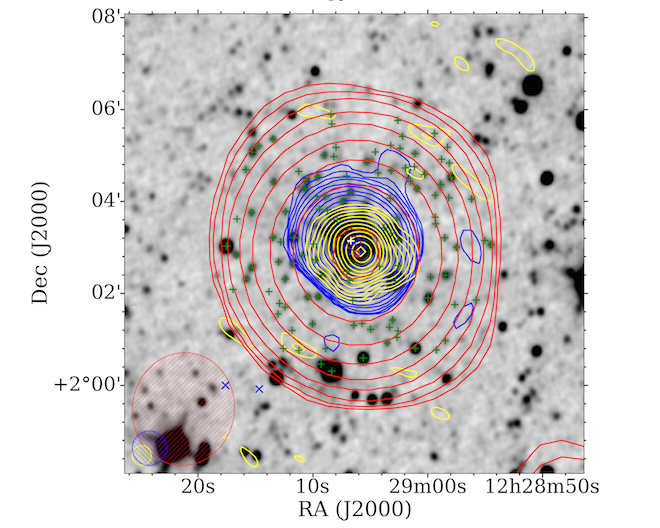}
	} 
\caption{Overlays for a few of the G4Jy sources that are subject to further checks against the literature (Section~\ref{sec:ambiguous}). The datasets, contours, symbols, and beams are the same as those used for Figure~\ref{FlameNebulaImage}, but where blue contours, crosses, and ellipses correspond to NVSS {\it or} SUMSS. In addition, positions from AllWISE are indicated by green plus signs, with that corresponding to the host galaxy highlighted in white. }
\label{furtherworkoverlays}
\end{figure*}

A radio map at 4.9\,GHz \citep{Bondi1993} allows us to identify the host galaxy for {\bf G4Jy~224} (GLEAM~J020704+293059). This source is 3C~59 (4C~+29.06), with very asymmetric morphology (Figure~\ref{furtherworkoverlays}a) and a pronounced hotspot where the eastern lobe terminates.

\begin{figure*}
\centering
\subfigure[G4Jy overlay for G4Jy~285]{
	\includegraphics[scale=1.1]{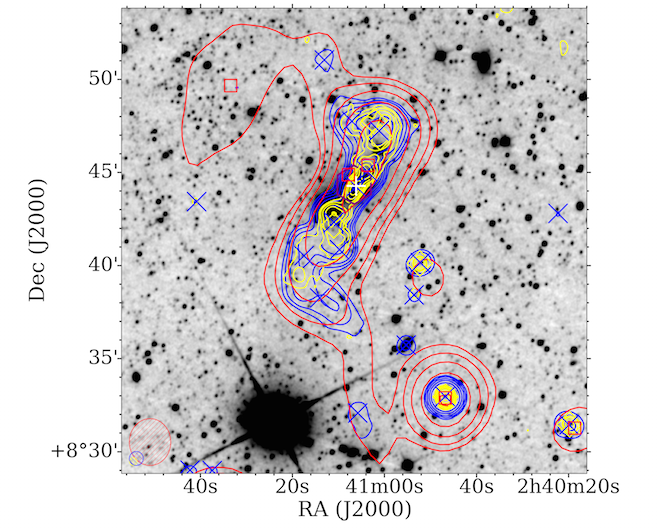}
	}
\subfigure[1.4-GHz/optical overlay for G4Jy~285]{
	\includegraphics[scale=1.1]{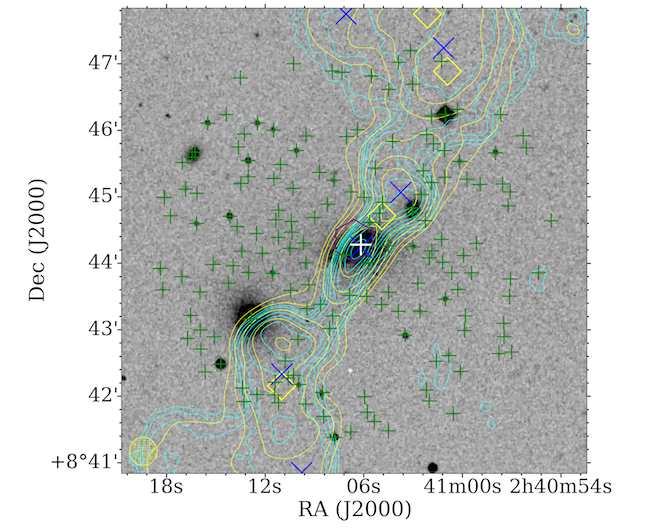}
	} 
\caption{Two overlays for G4Jy~285, which is NGC~1044 (Section~\ref{sec:ambiguous}). The first overlay, (a), uses the same datasets, contours, symbols, and beams as those used for Figure~\ref{FlameNebulaImage}. The second overlay, (b), uses an optical image from SuperCOSMOS (inverted greyscale) and a 1.4-GHz image (cyan contours) from \citet{Croston2008}, provided courtesy of Judith Croston. TGSS contours (in yellow) are shown for both overlays, and positions from AllWISE are indicated by green plus signs, with that corresponding to the host galaxy highlighted in white. }
\label{emailedCroston}
\end{figure*}

{\bf G4Jy~285} (GLEAM~J024103+084523 and GLEAM~J024107+084452) is NGC~1044 (4C +08.11). We use a 1.4-GHz radio image by \citet{Croston2008} to confirm that we have correctly identified the host galaxy (Figure~\ref{emailedCroston}), which is in agreement with \citet{vanVelzen2012}. We also note the interesting `zig-zag' of emission seen in the southern jet/lobe, and how the GLEAM contours trace trailing emission towards both the south and the north. In particular, there is fainter low-frequency emission (GLEAM~J024133+084940) towards the north-east. It is unclear whether or not this is associated with G4Jy~285, so we do not consider it further. 

{\bf G4Jy~287} (GLEAM~J024236-420133) appears in the Molonglo Southern 4-Jy sample (MS4; \citealt{Burgess2006a}) as B0240$-$422, and we agree with \citet{Burgess2006b} as to the host-galaxy identification. We therefore update the chosen AllWISE source to the one lying roughly midway between the two TGSS detections.

TGSS indicates that {\bf G4Jy~298} (GLEAM~J025102+061207) is a double, and we update the AllWISE position appropriately.

There are two good mid-infrared candidates for the host galaxy of {\bf G4Jy~311} (GLEAM~J025552$-$202743), which is PKS~B0253$-$206. \citet{ReidR1999} present an image at 4.7\,GHz, but it only shows the terminals of two lobes, and not the radio core. Being unable to resolve the ambiguity, we set the host flag to `u'.

We cannot find a high-resolution radio-image of {\bf G4Jy~318} (GLEAM~J030115$-$250538 and GLEAM~J030127$-$250354) in the literature, and so are uncertain as to the host-galaxy identification (figure~8c of Paper~I). Hence, we use `u' for the host flag.  

{\bf G4Jy~338} (GLEAM~J031605$-$265906) coincides with a crowded field in the mid-infrared. We could find no confirmation of the host in the literature, so set the host flag to `u'.

There is a trail of mid-infrared sources along the radio axis for {\bf G4Jy~350} (GLEAM~J032318$-$881613). With no high-resolution radio-image to discriminate between them, we use `u' for the host flag. 

Like for G4Jy~311, the image of {\bf G4Jy~376} (GLEAM~J034048$-$340851) by \citet{ReidR1999} shows the hotspots but no detection of the core$^{\ref{rkp1999note}}$. We leave the host flag as `u', as there are two good candidates for the host galaxy in the AllWISE image.

{\bf G4Jy~392} (GLEAM~J035231$-$071103) is 3C~94, with \citet{ReidR1999} confirming$^{\ref{rkp1999note}}$ that the AT20G detection is of a hotspot (belonging to a `double') and not of a core (belonging to a head-tail galaxy). We therefore update the host-galaxy position to that of the bright mid-infrared source located between the two NVSS detections (Figure~\ref{furtherworkoverlays}b).

The AllWISE source at the centroid position, and the one slightly south of the TGSS position, are both good host-galaxy candidates for {\bf G4Jy~413} (GLEAM~J040700$-$315210). This is PKS~B0405$-$320, but no high-resolution image is available so that we can check the core position.

{\bf G4Jy~414} (GLEAM~J040712+034318 and GLEAM~J040724+034049) has several candidate host-galaxies in the mid-infrared image. However, it is also known as 3C~105, which leads us to previous observations of the core \citep{Baum1988,Leahy1997}. These allow us to manually select the correct host-galaxy, and so update the mid-infrared position for this source. 

{\bf G4Jy~434} (GLEAM~J041508$-$292901) is PKS~B0413$-$296, and a 4.9-GHz image by \citet{Kapahi1998} shows that this is a very asymmetric `double'. We update our host-galaxy position, selecting the AllWISE source that is consistent with the core position.

{\bf G4Jy~441} (GLEAM~J042020+151659) is 4C~+15.13, but no high-resolution image is available. Extension in the TGSS contours could be because this is a head-tail galaxy, in which case the AllWISE source coincident with the TGSS position is the host. Alternatively, G4Jy~441 could be a `double', with a host galaxy lying between the two NVSS detections but too faint to appear in AllWISE. Hence, we set the host flag to `u'.

A 1.4-GHz image by \citet{Kapahi1998} allows us to confirm that the host galaxy of {\bf G4Jy~448} (GLEAM~J042234$-$261700; PKS~B0420$-$26) is the AllWISE source closest to the centroid position.

{\bf G4Jy~488} (GLEAM~J044111+251845) is 4C~+25.15, but we did not find a high-resolution radio image in the literature. Although AllWISE~J044111.01+251839.6 is the likely host, it is blended with another AllWISE source, so we use `u' for the host flag.

A 4.8-GHz image by \citet{Antonucci1985}, and a core position quoted by \citet{Ekers1989}, allow us to confirm that the central 6dFGS source (g0448306$-$203214) is the host of {\bf G4Jy~499} (GLEAM~J044829$-$203217; PKS~B0446$-$20). It is unclear how much of the extended emission is associated with the `double' (Figure~\ref{furtherworkoverlays}c), so we leave the morphology label as `complex'. 

{\bf G4Jy~530} (GLEAM~J051250$-$482358) is PKS~B0511$-$48. Whilst \citet{Smith1985} are likely correct in their optical identification, their radio image is not of sufficient resolution to rule out another mid-infrared source that is also on-axis. We could not find a better image in the literature and so set the host flag to `u'.

For {\bf G4Jy~570} (GLEAM~J054049$-$614233; B0540$-$617), the nearest AllWISE detection to the centroid appears to be incorrect. Given the `pinching' of the radio contours associated with the southern lobe, we are happy to update the host-galaxy position so that it is in agreement with the optical identification provided by \citet{Jones1992}. 

{\bf G4Jy~580} (GLEAM~J054924$-$405110) is in the MS4 sample \citep{Burgess2006a,Burgess2006b}, and we agree with their identification. The latter is supported by an ATCA image, indicating that the AT20G detection corresponds to a hotspot in the north-west lobe.

The extended radio-galaxy {\bf G4Jy~680} (GLEAM~J080225$-$095823, GLEAM~J080253$-$095822) has several mid-infrared sources that could be interpreted as the host galaxy (Figure~\ref{furtherworkoverlays}d). Previous identifications for this source -- referred to as B0800$-$09 -- focus on optical data at the two positions indicated by 6dFGS detections (immediately east and west of the centroid). Namely, \citet{Danziger1983} quote both the identification favoured by \citet{Schilizzi1975b} -- g0802363$-$095740 in 6dFGS, at $z = 0.0699$ -- and the alternative host-galaxy at $z = 0.0865$ (g0802400$-$095735 in 6dFGS, at $z = 0.0858$). There is no obvious core in NVSS, nor coverage by FIRST, which may have allowed us to resolve the ambiguity, or indeed reveal that an {\it obscured} galaxy is the true host. As such, we retain the `u' host-flag, as originally assigned. 

A FIRST image suggests that {\bf G4Jy~727} (GLEAM~J084356+154738) is a compact `double', with a host position that is offset from the nearest AllWISE source. As the host is not seen in the mid-infrared image, we set the host flag to `m'.

We use a 4.9-GHz image \citep{Morganti1993} to confirm the host galaxy for {\bf G4Jy~747} (GLEAM~J090147$-$255516; B0859$-$25). Their image also indicates that the AT20G detection (seen in our overlay) is the result of a hotspot.

{\bf G4Jy~767} (GLEAM~J092158+082850) and {\bf G4Jy~810} (GLEAM~J100028+140134) both have a `double' morphology in FIRST, with these FIRST images supporting our host-galaxy identifications.

\begin{figure}
\centering
\includegraphics[scale=1.1]{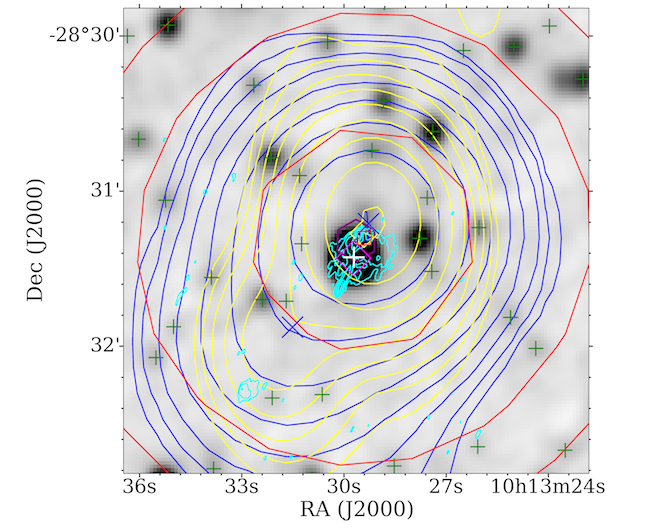}
\caption{An overlay for G4Jy~829 (Section~\ref{sec:ambiguous}), where the datasets, contours, and symbols are the same as those used for Figure~\ref{FlameNebulaImage}. In addition, positions from AllWISE are indicated by green plus signs (with the host galaxy highlighted in white) and the cyan contours are from a 4.7-GHz image \citep{ReidR1999}. }
\label{doubledoubleat5GHz}
\end{figure}

At 4.7\,GHz \citep{ReidR1999}$^{\ref{rkp1999note}}$ we see that the 6dFGS source, g1013297$-$283126, is indeed the host galaxy of {\bf G4Jy~829} (GLEAM~J101329$-$283118). This source is also known as PKS~B1011$-$282, and we suggest that it is a `double-double' radio-galaxy, with inner jets and only the southern outer-lobe seen in the 4.7-GHz image (Figure~\ref{doubledoubleat5GHz}). The northern outer-lobe may be resolved out, or simply too faint to be detected at this high frequency.

{\bf G4Jy~844} (GLEAM~J102529$-$021739) has two host-galaxy candidates very close to the centroid position. However, this source has a `double' morphology in FIRST, which allows us to select the western AllWISE source. 

Similarly, `triple' morphology in FIRST allows us to confirm the host galaxy for {\bf G4Jy~884} (GLEAM~J105817+195203).

Further investigation was also carried out for {\bf G4Jy~886} (GLEAM~J105846$-$361754 and GLEAM~J105854$-$362051), in order to check that the 6dFGS source close to the centroid (Figure~\ref{furtherworkoverlays}e) is the correct host-galaxy. The position for this optical identification (g1058548$-$361921 at $z = 0.0705$) is confirmed by the coincidence of the elliptical galaxy with a weak radio-core (B1056$-$360 in \citealt{Ekers1989}). 

FIRST shows that {\bf G4Jy~888} (GLEAM~J110203$-$011619) has a `double' morphology, allowing us to confirm that the AllWISE source closest to the centroid position is the host galaxy. The AT20G detection coincides with a neighbouring AllWISE source, so it likely marks either the core of an unrelated radio source, or a hotspot belonging to G4Jy~888.

{\bf G4Jy~917} (GLEAM~J112554$-$352321; B1123$-$351) has a dense field of mid-infrared sources, this being cluster AS~665. We confirm the position of the radio core based upon a 4.9-GHz image presented by \citet{Ekers1989}. As a result, we agree that the host galaxy of this radio source is ESO~377$-$G046 (g1125529$-$352340), and that the identification by \citet{vanVelzen2012} is consistent with the original optical identification \citep{Bolton1965}. 

For {\bf G4Jy~934} (GLEAM~J113917$-$322237; PKS~B1136$-$32) we dismiss the AllWISE source that is coincident with the centroid position, and instead select the AllWISE source that is further north along the radio axis. This host-galaxy identification is in agreement with the MS4 identification \citep{Burgess2006b}, which itself is based upon imaging of the possible core \citep{Duncan1992}. 

We set `u' as the host flag for {\bf G4Jy~939} (GLEAM~J114134$-$285050) as we are unable to find a high-resolution image to distinguish between multiple mid-infrared candidates close to the centroid. 

\begin{figure*}
\centering
\subfigure[1\arcdeg\ overlay for G4Jy~990]{	\includegraphics[scale=1.1]{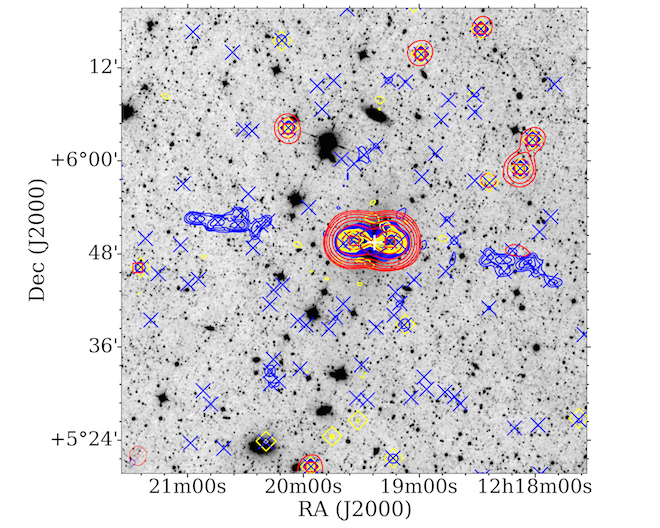}
	}
\subfigure[12$'$ overlay for G4Jy~990]{	\includegraphics[scale=1.1]{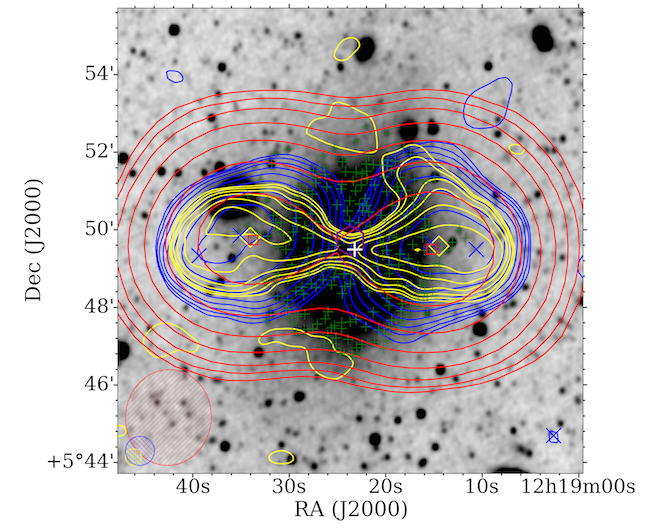}
	}
\caption{Two overlays for G4Jy~990, using the same datasets, contours, symbols, and beams as those used for Figure~\ref{FlameNebulaImage}. In (a), the mirrored, blue contours either side of the source are believed to be artefacts in NVSS (Section~\ref{sec:ambiguous}). In (b), positions from AllWISE are indicated by green plus signs, with that corresponding to the host galaxy highlighted in white.}
\label{mirroredNVSSartefacts}
\end{figure*}

The 1\arcdeg\ overlay for {\bf G4Jy~990} (GLEAM~J121915+054929, GLEAM~J121933+054944) shows mirrored, extended, NVSS emission, which we deem to be artefacts (Figure~\ref{mirroredNVSSartefacts}a). This source is 3C~270 (NGC~4261), for which our AllWISE identification (Figure~\ref{mirroredNVSSartefacts}b) is in agreement with the core position \citep{Morganti1993}.

{\bf G4Jy~1001} (GLEAM~J122811+202321) shows `triple' morphology in FIRST, and we update the AllWISE position to that coincident with the radio core. 

The AllWISE source nearest to the centroid position for {\bf G4Jy~1003} (GLEAM~J122906+020251) may appear to be a star -- given the strong diffraction-spikes in the mid-infrared image (Figure~\ref{furtherworkoverlays}f) -- but it is actually the optically-brightest quasar in the sky, 3C~273. This AllWISE position is in agreement with abundant literature (see NED) as to the radio-core/host-galaxy location.

{\bf G4Jy~1021} (GLEAM~J124602+255359 and GLEAM~J124612+255337) shows `triple' morphology in FIRST. As such, despite the density of mid-infrared sources (figure~8f of Paper~I), we are able to identify the host galaxy.  

\begin{figure*}
\centering
\subfigure[G4Jy overlay for G4Jy~1040]{
	\includegraphics[scale=1.1]{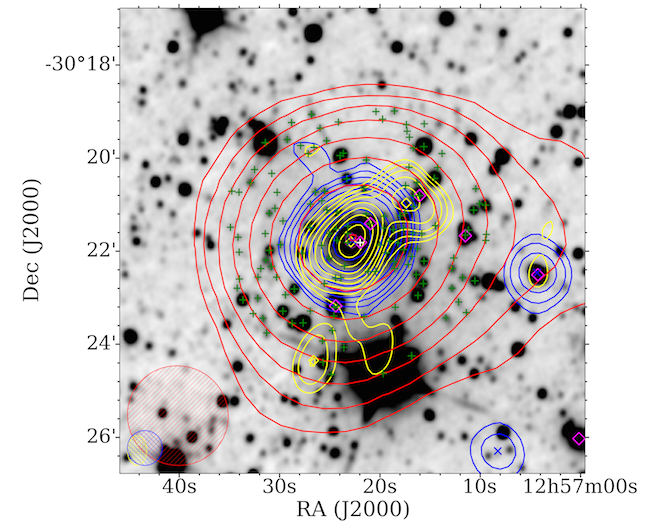}
	}
\subfigure[2.4-GHz contours for G4Jy~1040]{
	\includegraphics[scale=1.1]{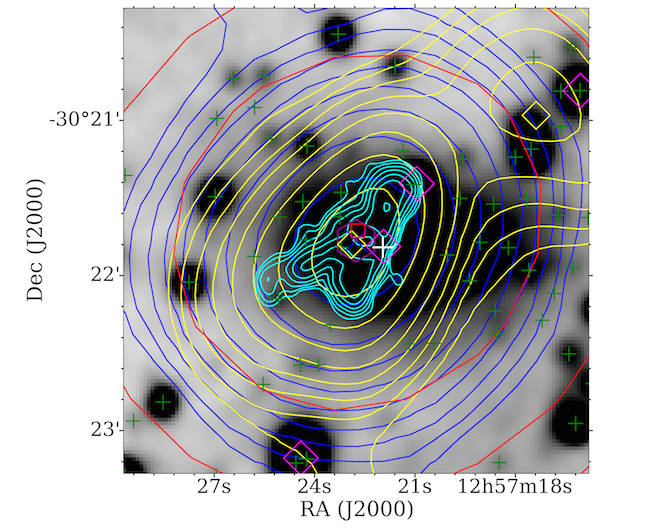}
	} 
\caption{Two overlays for G4Jy~1040 (Section~\ref{sec:ambiguous}), the first of which, (a), uses the same datasets, contours, symbols, and beams as those used for Figure~\ref{FlameNebulaImage}. The second overlay, (b), is a zoomed-in version that also uses a 2.4-GHz image (\citealt{Venturi2001}; cyan contours) provided courtesy of Tiziana Venturi. Positions from AllWISE are indicated by green plus signs, with that corresponding to the host galaxy highlighted in white. }
\label{emailedVenturi}
\end{figure*}

{\bf G4Jy~1040} (GLEAM~J125722$-$302142) is in the cluster Abell~3532 (of the A3528-A3530-A3532 merging complex), and the 2.4-GHz image provided by \citet{Venturi2001} allows us to confirm that g1257219$-$302149 is the host galaxy (Figure~\ref{emailedVenturi}).

FIRST reveals a `double' morphology for {\bf G4Jy~1099} (GLEAM~J135209$-$054345), and so we can confirm that we have correctly identified the host galaxy in AllWISE. 

For {\bf G4Jy~1158} (GLEAM~J142432$-$491344; PKS~B1421$-$490), there are two candidate mid-infrared hosts close to the centroid position. We resolve this ambiguity by consulting \citet{Godfrey2009}, whose 20.2-GHz image confirms the position of the core, and therefore the host galaxy. Their image also suggests that the AT20G detection marks a hotspot.

We set `m' as the host flag for {\bf G4Jy~1160} (GLEAM~J142457+200021) as the nearest AllWISE detection appears to be a blend in the mid-infrared image, with a 4$''$ offset between the AllWISE position and the coincident radio-positions from multiple surveys. A similar scenario is seen for {\bf G4Jy~1167} (GLEAM~J142740+283327), but in this case the offset is 7$''$. Again, the host flag is `m'.

{\bf G4Jy~1180} (GLEAM~J143719$-$075339) is PKS~B1434$-$076. A 4.7-GHz image \citep{ReidR1999} shows `triple' morphology$^{\ref{rkp1999note}}$, and confirms that the AllWISE position between the two TGSS detections is the host of the radio emission. 

{\bf G4Jy~1190} (GLEAM~J144635$-$084544) is PKS~B1443$-$085 in cluster Abell~1964. We consult \citet{Owen1992} but the morphology and host galaxy are still unclear in their 1.5-GHz image. We therefore assign the source the `complex' morphology-label and `u' as the host flag. 

`Triple' morphology is seen at 4.7\,GHz \citep{ReidR1999} for {\bf G4Jy~1209} (GLEAM~J145555$-$110856; PKS~B1453$-$10). This allows us to update our host-galaxy identification, which is in agreement with the core position$^{\ref{rkp1999note}}$. The image also indicates that AT20G has detected a hotspot in the southern lobe.

\begin{figure*}
\centering
\subfigure[G4Jy~1216]{
	\includegraphics[scale=1.1]{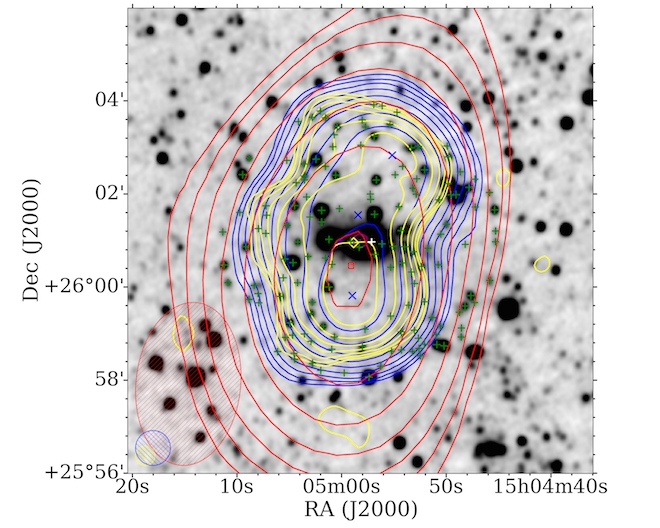}
	} 
\subfigure[G4Jy~1265]{
	\includegraphics[scale=1.1]{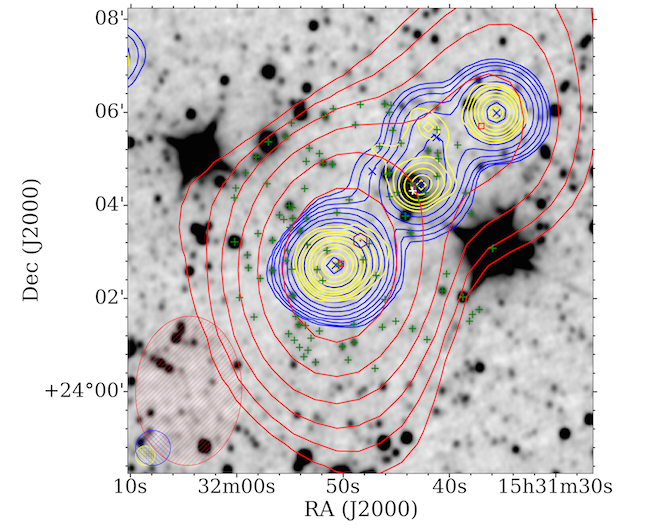}
	} 
\caption{Overlays for two more G4Jy sources that are subject to further checks against the literature (Section~\ref{sec:ambiguous}). The datasets, contours, symbols, and beams are the same as those used for Figure~\ref{FlameNebulaImage}. In addition, positions from AllWISE are indicated by green plus signs, with that corresponding to the host galaxy highlighted in white. }
\label{furtherworkoverlays2}
\end{figure*}

There are several mid-infrared sources close to the centroid position for {\bf G4Jy~1216} (GLEAM~J150459+260027; Figure~\ref{furtherworkoverlays2}a), and so we originally assigned this object `u' for the host flag. However, previous observations of this source (also known as 3C~310 in 3CRR) allow us to identify the correct mid-infrared host. That is, \citet{Burns1979} show that the radio core coincides with the western source of a galaxy pair, at $z=0.054$. Therefore, the lobe-to-lobe separation of 5$\farcm$1 corresponds to a physical size of 322\,kpc.

In the case of {\bf G4Jy~1265} (GLEAM~J153137+240542 and GLEAM~J153150+240244), the candidate host-galaxies are relatively far from the centroid position (Figure~\ref{furtherworkoverlays2}b). As to which one is the correct cross-identification, we note that this radio source (3C~321 in 3CRR) has an optical position provided by \citet{Jenkins1977}. This is spectroscopically confirmed by \citet[][]{deGrijp1992}, and the acquired redshift of $z=0.096$ means that G4Jy~1265 (spanning 5$\farcm$1 in angular size) extends over 544\,kpc.

{\bf G4Jy~1323} (GLEAM~J162021+173623) is in the 3CRR sample, as 3C~334, and has `triple' morphology at the resolution provided by FIRST. The bright mid-infrared source near the centroid is its host galaxy, as confirmed by the location of the core.

We set the host flag for {\bf G4Jy~1324} (GLEAM~J162033$-$710017) to `u', as there are two good candidates for the host galaxy but no high-resolution image to confirm the correct one. 

{\bf G4Jy~1332} (GLEAM~J162439+234512) has `double' morphology in FIRST that supports our host-galaxy identification.

A FIRST image confirms that {\bf G4Jy~1338} (GLEAM~J162732+211224) is a point source, unrelated to the faint `double' towards the north-west.

{\bf G4Jy~1365} (GLEAM~J164604$-$222756) is PKS~B1643$-$22 and it is unclear which of two AllWISE positions marks the host galaxy. We have checked against the literature but cannot find a high-resolution image to resolve the ambiguity (hence setting `u' as the host flag). The AT20G detection could be the radio core of the northern candidate, or a hotspot associated with the southern candidate.

For {\bf G4Jy~1377} (GLEAM~J165712$-$134911), the AllWISE position nearest to the centroid may be the host galaxy, but its offset from the radio axis prompts further investigation. It is PKS~B1654$-$137 and we find no high-resolution image in the literature. Due to the possibility that there is a faint mid-infrared host that is on-axis, we set the host flag to `u'.

{\bf G4Jy~1402} (GLEAM~J172031$-$005845) is 3C~353 (B1717$-$00). We use the image presented by \citet{Morganti1993} to confirm that the 6dFGS source is the host galaxy. 

{\bf G4Jy~1438} (GLEAM~J174630+234252) is 4C~+23.45. Although AllWISE~J174631.15+234325.9 is likely the host galaxy, mid-way between the two NVSS detections, we have no better-resolution radio image to confirm this. The host flag is therefore set to `u'.

Our host-galaxy identification for {\bf G4Jy~1478} (GLEAM~J182220$-$554148) is updated to the AllWISE source mid-way between the two SUMSS components of this `double'. This source is also known as PKS~B1818$-$557, and our selection is in agreement with \citet{Burgess2006b}.

It is unclear whether {\bf G4Jy~1491} (GLEAM~J183356$-$394023; PKS~B1830$-$39) has head-tail, `double' or core-jet morphology. With no better-resolution radio image available in the literature, we label the morphology `complex' and use `u' for the host flag. 

{\bf G4Jy~1499} (GLEAM~J183825+171154) is 3C~386 in the 3CRR sample. We use an image of the core presented by \citet{Leahy1991} to confirm our identification in AllWISE.  

{\bf G4Jy~1537} (GLEAM~J192606$-$573954) is B1921$-$577. The nearest AllWISE source to the centroid is AllWISE~J192608.47$-$574004.1, which is the $b_J = 19$\,mag galaxy mentioned by \citet{Jones1992}. However, they also note a $b_J = 16.5$\,mag galaxy further to the west, and observations at 1.4\,GHz \citep{Jones1992b} show that the radio lobes are emanating from this position. Therefore, we update the mid-infrared identification accordingly (to AllWISE~J192605.75$-$574016.4), and note that it coincides with a detection in 6dFGS.

\begin{figure*}
\centering
\subfigure[ G4Jy overlay for G4Jy~1539]{
        \includegraphics[scale=1.1]{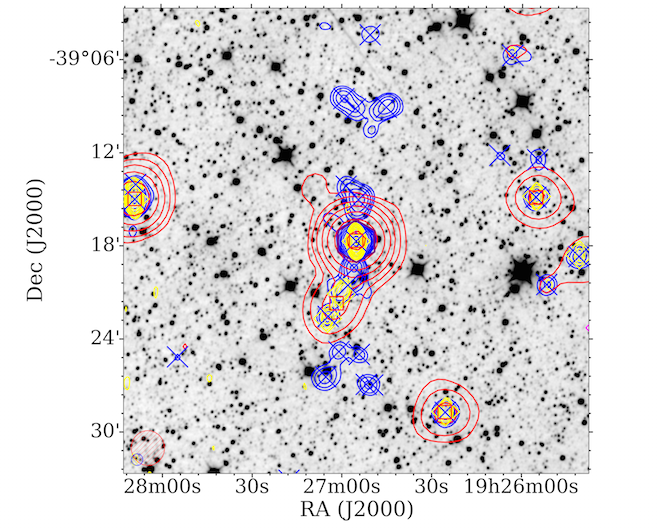}
	}
\subfigure[NVSS image for G4Jy~1539]{
	\includegraphics[scale=1.1]{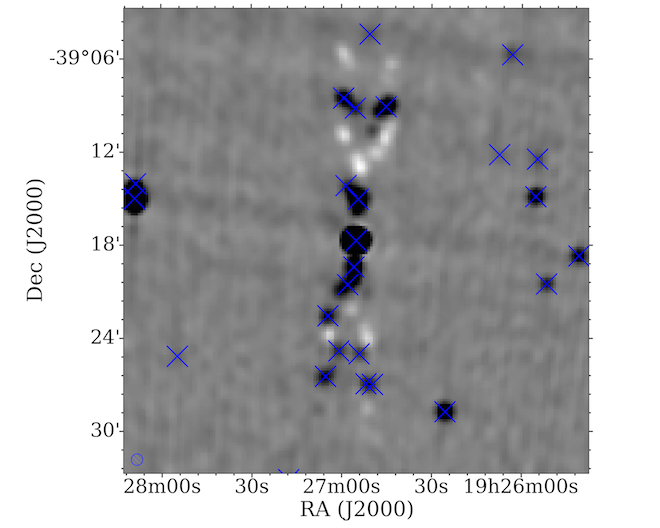}
	}
\caption{G4Jy~1539 (Section~\ref{sec:ambiguous}), as it appears (a) in the overlay used for visual inspection, and (b) in NVSS (inverted greyscale). The overlay uses the same datasets, contours, symbols, and beams as those used for Figure~\ref{FlameNebulaImage}. \label{chromosome}}
\end{figure*}

It is unclear whether the host for {\bf G4Jy~1539} (GLEAM~J192655$-$391741) is significantly offset from the centroid position, or too faint to be seen in the mid-infrared image. We therefore set the host flag to `u'. For the morphology label we use `complex', as we cannot determine whether nearby extended emission is associated with the source or connected to artefacts in NVSS. Due to some of these artefacts having negative flux-density, their unusual chromosome-like morphology is only seen in the full NVSS image rather than the contours (Figure~\ref{chromosome}).

{\bf G4Jy~1540} (GLEAM~J192724$-$225842) has two host-galaxy candidates close to the centroid that are both approximately on the radio axis. We check this source against the literature but cannot find a radio image to resolve the ambiguity, so leave the host flag as `u'.

An ATCA image by \citet{Burgess2006b} indicates that the AT20G detection associated with {\bf G4Jy~1558} (GLEAM~J193557$-$462040; B1932$-$464) is due to a hotspot, and not the core. We update the AllWISE host-galaxy accordingly, where it is now further from the centroid and in agreement with the MS4 identification.

{\bf G4Jy~1590} (GLEAM~J195817$-$550934) is B1954$-$55, for which \citet{Morganti1999} present a 8.6-GHz ATCA image that clearly distinguishes two lobes and a core (i.e. `triple' morphology). The nearest AllWISE position (AllWISE~J195816.66$-$550934.9) to the core is offset by $\sim3''$ and, as we discover via SIMBAD, is a star (UCAC2 7908216). As this star is obscuring the mid-infrared host of G4Jy~1590, we set the host flag to `m'. 

{\bf G4Jy~1617} (GLEAM~J202336+170057 and GLEAM~J202343+170549) is 4C~+16.68. There is no high-resolution radio-image in the literature to confirm the host galaxy, so we leave the host flag as `u'.

We use 1.4-GHz observations by \citet{Worrall2014} to confirm the host galaxies for {\bf G4Jy~1677} and {\bf G4Jy~1678}, which were re-fitted for the G4Jy catalogue (appendix~D.3 of Paper~I). They are both in cluster Abell~3744, and also known as NGC~7016 and NGC~7018, respectively. At the high resolution provided by the VLA in `A'-array configuration, the northern jet of G4Jy~1677 is seen to loop back onto itself. Meanwhile, observations in `C'-array configuration \citep{Bicknell1990} allow more-extended emission to be detected, which reveals long `tendrils' associated with both G4Jy~1677 and G4Jy~1678. The tendrils of the northern source are also seen in the GLEAM contours, thanks to the MWA's sensitivity to diffuse emission. 

{\bf G4Jy~1694} (GLEAM~J212616$-$552112) appears in the MS4 sample as B2122$-$555, with `triple' morphology seen in an ATCA image \citep{Burgess2006b}. The AT20G position in our overlay coincides with the northern lobe, implying that it marks a hotspot. Meanwhile, the core position allows us to distinguish between two host-galaxy candidates in AllWISE that lie between the two SUMSS positions. As `double' morphology is not clear in the SUMSS contours, we label G4Jy~1694 as `single'.

The `double' radio-galaxy, {\bf G4Jy~1773} (GLEAM~J222106$-$501818), is B2217$-$505. We update the AllWISE identification to a source that is slightly further from the centroid position, but coincident with the TGSS detection and on the SUMSS radio-axis. This source is also consistent with the optical identification provided by \citet{Jones1992}.

Another MS4 source is {\bf G4Jy~1795} (GLEAM~J225303$-$405744; B2250$-$412), which shows `double' morphology when observed with ATCA \citep{Burgess2006b}. Again, AT20G marks a hotspot, and our AllWISE host-galaxy is in agreement with the MS4 identification.

\subsection{Possible disagreement with the existing host-galaxy identification}
\label{sec:disagreement}

\begin{figure*}
\centering
\subfigure[G4Jy~453]{
	\includegraphics[scale=1.1]{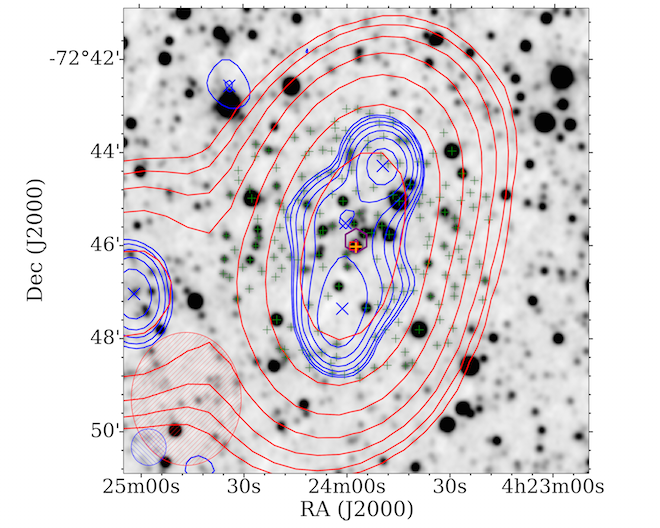}
	}
\subfigure[G4Jy~641]{
	\includegraphics[scale=1.1]{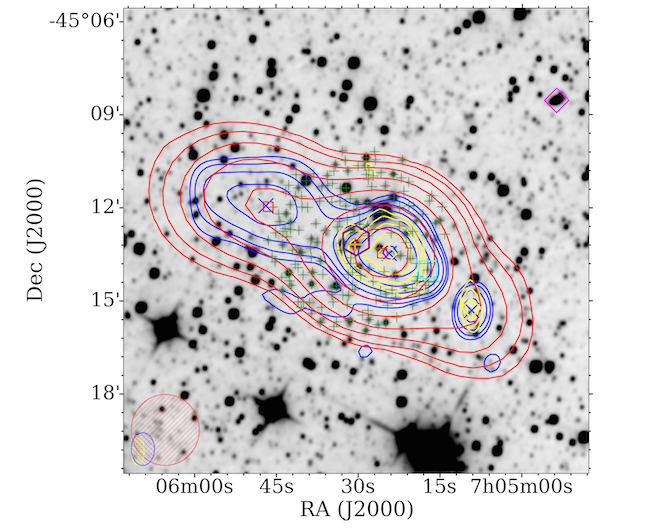}
	}
\subfigure[G4Jy~700]{
	\includegraphics[scale=1.1]{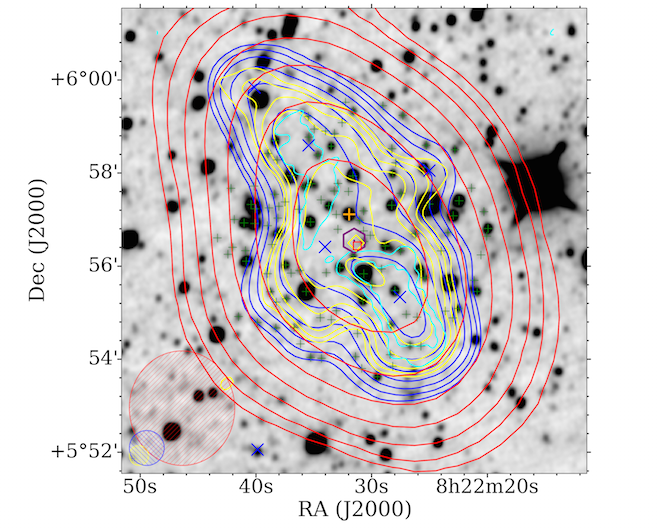}
	}
\subfigure[G4Jy~1582]{
	\includegraphics[scale=1.1]{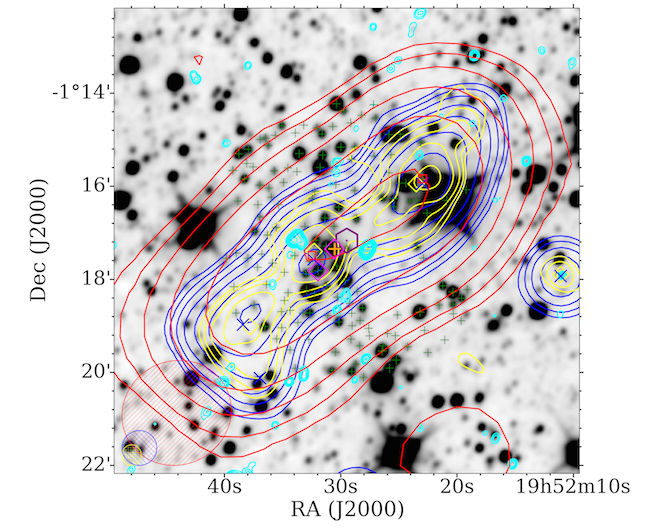}
	}
\caption{Overlays for G4Jy sources for which there is doubt (Section~\ref{sec:disagreement}) over the existing host-galaxy identification in the literature (orange plus signs). The datasets, contours, symbols, and beams are the same as those used for Figure~\ref{FlameNebulaImage}, but where blue contours, crosses, and ellipses correspond to NVSS {\it or} SUMSS. AllWISE positions within 3$'$ of the centroid (purple hexagon) are indicated by green plus signs, and for G4Jy~700 and G4Jy~1582 we plot (in cyan) 4.9-GHz and 300-MHz contours, respectively, using VLA images provided courtesy of Francesco Massaro \citep{Massaro2012}.}
\label{disagreementoverlays}
\end{figure*}


{\bf G4Jy~453} (GLEAM~J042358$-$724601) appears in the literature as B0424$-$728, and the nearest mid-infrared source to its centroid (AllWISE~J042357.45$-$724602.0) is coincident with the optical identification noted by \citet{Jones1992}. The interpretation of this source as a GRG \citep[e.g.][]{Subrahmanyan1996} relies on the corresponding redshift. However, a third SUMSS detection suggests that the core of this radio galaxy may actually be at/near R.A. = 04:24:00.42, Dec. = $-$72:45:31.8 (J2000), north-east of the previous identification (see Figure~\ref{disagreementoverlays}a). Due to this inconsistency, and the density of mid-infrared candidates, we are uncertain as to the true host-galaxy position and so specify `u' for the host flag.

{\bf G4Jy~641} (GLEAM~J070525$-$451328 and GLEAM~J070546$-$451158) is B0703$-$451. This source is also believed to be a GRG, although \citet{Malarecki2015} acknowledge that `the host galaxy of B0703$-$451 is in doubt'. We can see in Figure~\ref{disagreementoverlays}b why this is the case, with multiple mid-infrared candidates for the host. The current identification is AllWISE~J070530.57$-$451311.1, which is the AllWISE source nearest to the centroid position. This is in agreement with the optical identification by \citet{Jones1992} but a high-resolution radio-image to confirm this is unavailable. Hence, we do not list this source in Table~\ref{listofGRGs}, and instead set the host flag to `u', for prompting follow-up observations.

{\bf G4Jy~700} (GLEAM~J082231+055626) can be found in the 3CR catalogue, where it is listed as 3C~198. It is in cluster Abell~115, and the cross-identification provided is a $V = 17$\,mag galaxy \citep{Wyndham1966} corresponding to AllWISE~J082231.95+055706.8. However, we are wary of historical identifications that may be biased towards the optically-brightest galaxies, hence our use of mid-infrared images (which also show dust-obscured galaxies). During our reassessment of this particular source, we find 4.9-GHz radio contours in the literature, presented by \citet{Massaro2012}. They are unable to locate the core in the X-ray, and nor is it detected in the radio. However, the radio image reveals that the lobes of this extended radio galaxy are bent (Figure~\ref{disagreementoverlays}c). This could be due to backflow of plasma towards the south-east, in a similar way as seen for 3C~28\footnote{http://www.jb.man.ac.uk/atlas/object/3C28.html, as part of the `Atlas of DRAGNs' database assembled by \citet{Leahy1996}.} (G4Jy~99; GLEAM~J005550+262436). Alternatively, the narrowing of the radio emission could be the {\it inner} parts of the lobes, with the host galaxy being nearby and therefore not in line with the major axis of the radio galaxy. If so, this could explain why \citet{Buttiglione2010} -- who use the existing identification -- find 3C~198 to be the only object (out of 113 3CR sources considered) to have optical emission-lines consistent with star-forming galaxies rather than AGN. In any case, and in keeping with other extended radio-galaxies in our sample that have ambiguous hosts, we use `u' for the host flag until further evidence becomes available.

Another 3CR source for which we question the existing host-galaxy identification is 3C~403.1. This appears in the G4Jy Sample as {\bf G4Jy~1582} (GLEAM~J195222$-$011550 and GLEAM~J195232$-$011729), with its current identification coinciding with AllWISE~J195230.53$-$011720.7 and g1952305$-$011721 (in 6dFGS). There is weak (0.5--7\,keV) X-ray emission at this position \citep{Massaro2012}, but the same can be said for the nearby source, AllWISE~J195231.25$-$011716.6. Hence, there are two AGN candidates for G4Jy~1582, and the VLA image at 300\,MHz does not allow this ambiguity to be resolved. However, interestingly, the orientation of the radio axis at 300\,MHz (cyan contours in Figure~\ref{disagreementoverlays}d) appears to be at $\sim45$\arcdeg\ with respect to the radio axis at 150\,MHz (yellow contours in Figure~\ref{disagreementoverlays}d), and so is suggestive of radio-jet precession (see Section~\ref{sec:XandSorZshaped}).


Like the `unclassifiable' sources described in Section~\ref{sec:unclassifiable}, each of the above four sources are being followed up using MeerKAT (PI: White).

\section{Summary} \label{sec:finalsummary}

The multi-component nature and complex morphology of many radio sources means that their host galaxies need to be carefully identified before we can combine radio datasets with information at other wavelengths. We have completed this important step for the brightest radio-sources ($S_{\mathrm{151\,MHz}}>4$\,Jy) in the GLEAM extragalactic catalogue (EGC), which we refer to as the GLEAM 4-Jy (G4Jy) Sample. Here we summarise the work done as part of cross-identifying this sample, and in doing so, we provide added value through the G4Jy catalogue (see Paper~I for details):

\begin{enumerate}

\item{We visually inspect each of the G4Jy sources (as defined in Paper~I) by overlaying multiple sets of radio contours onto mid-infrared images, with candidate host-galaxies marked by AllWISE catalogue positions. All of the overlays that we use for this work, as well as the images from which they are made, are accessible online (see https://github.com/svw26/G4Jy for details).}

\item{Based upon our visual inspection and numerous literature checks, we highlight the wide variety of bright radio-sources in the sample. This includes 2 nearby, star-forming galaxies, 8 known GRGs, and 14 S-/Z-/X-shaped sources. We also create lists of 23 bent-tail radio-galaxies and 18 head-tail radio-galaxies (subject to the resolution of our radio images), whose morphology may be used to probe the surrounding cluster medium.}

\item{For 1,606 of the 1,863 sources in the G4Jy Sample, we are able to identify the host galaxy in AllWISE, supported by higher-resolution (i.e. $<$\,25--45$''$) images in the literature, where available. Meanwhile, another 126 sources are deemed to have a host galaxy that is too faint to appear in the AllWISE catalogue (which includes those affected by nearby, bright, mid-infrared emission). Many of these sources are likely to be at high redshift, and require follow-up to confirm this. The two radio sources for which a host-galaxy identification is inappropriate (and so not specified) is the Flame Nebula (G4Jy~571) and the northern cluster-relic of Abell~3667 (G4Jy~1605).}

\item{We flag the remaining 129 G4Jy sources as having a cross-identification that is `uncertain' (i.e. host\_flag = `u'). This includes our wariness over existing identifications for B0424$-$728 (G4Jy~453), B0703$-$451 (G4Jy~641), 3C~198 (G4Jy~700) and 3C~403.1 (G4Jy~1582), as we do not find sufficient evidence for the current host-galaxy position. In addition, there are three sources (G4Jy~113, G4Jy~513, and G4Jy~1117) for which we are unable to infer the mechanism giving rise to the low-frequency emission, and therefore it is unclear whether or not a host galaxy should be identified. Hence, each of these 129 sources are being followed up with MeerKAT (PI: White) to aid investigation.}

\item{Our fresh assessment of these bright radio-sources, coupled with attention to detail, allow us to identify (i) discrepancies in the literature regarding the interpretation of morphology (e.g. for G4Jy~325, also known as B0304$-$123A), and (ii) inconsistent co-ordinates for existing high-resolution radio-images (e.g. for G4Jy~40; PKS~B0018$-$19).}

\item{Of course, our interpretations and identifications are themselves limited by the spatial resolution of the radio images at hand. Whilst NVSS/SUMSS allows us to identify which sources likely have their low-frequency flux densities affected by confusion (relevant for $\sim$20\% of the sample), images of 45$''$ resolution may still be concealing: (i) multiple, unrelated sources, (ii) the presence of distinct/asymmetric radio jets/lobes (which includes core-jet systems), or (iii) bent-/head-tail morphology. It is expected, therefore, that many morphology labels provided in the G4Jy catalogue will need to be revised in the future.}

\end{enumerate}

Finally, for the sources being followed up with MeerKAT at $\sim$5$''$ resolution, we will use the same methods for host-galaxy identification (outlined in section~5.5 of Paper~I) and update the G4Jy catalogue accordingly.


\section{Dedication}

Papers I and II are dedicated to the memory of Richard Hunstead, who was very helpful with the assessment of the sources presented in this work, and provided hitherto unpublished radio-images.

\begin{acknowledgements}

We thank the anonymous referee for their careful review and very positive feedback. SVW would like to thank Chris Jordan, for his help with installing and running the cross-matching software (https://github.com/kasekun/MCVCM), as well as Dave Pallot and the ICRAR Data Intensive Astronomy team, for their help with fixing and updating the G4Jy Sample Server. In addition, SVW thanks Ron Ekers, Elaine Sadler, Guillaume Drouart and Gulay {G{\"u}rkan} for useful discussions, and Francesco Massaro, Judith Croston and Tiziana Venturi, for sharing images. We acknowledge the International Centre for Radio Astronomy Research (ICRAR), which is a joint venture between Curtin University and The University of Western Australia, funded by the Western Australian State government. We acknowledge the Pawsey Supercomputing Centre which is supported by the Western Australian and Australian Governments. The financial assistance of the South African Radio Astronomy Observatory (SARAO) towards this research is hereby acknowledged (www.ska.ac.za).

This scientific work makes use of the Murchison Radio-astronomy Observatory, operated by CSIRO. We acknowledge the Wajarri Yamatji people as the traditional owners of the Observatory site. Support for the operation of the MWA is provided by the Australian Government (NCRIS), under a contract to Curtin University administered by Astronomy Australia Limited. 

GMRT is run by the National Centre for Radio Astrophysics of the Tata Institute of Fundamental Research (TIFR). The Australia Telescope Compact Array (ATCA) is part of the Australia Telescope National Facility which is funded by the Australian Government for operation as a National Facility managed by CSIRO.

This publication made use of data products from the {\it Wide-field Infrared Survey Explorer}, which is a joint project of the University of California, Los Angeles, and the Jet Propulsion Laboratory/California Institute of Technology, funded by the National Aeronautics and Space Administration (NASA).

This research has made use of the NASA/IPAC Extragalactic Database (NED), which is operated by the Jet Propulsion Laboratory, California Institute of Technology, under contract with the National Aeronautics and Space Administration. We also acknowledge the use of NASA's SkyView facility [http://skyview.gsfc.nasa.gov] located at NASA Goddard Space Flight Center. This research has made use of the SIMBAD astronomical database \citep{Wenger2000}, operated at CDS, Strasbourg, France.

This research has made use of the VizieR catalogue access tool, CDS, Strasbourg, France (DOI: 10.26093/cds/vizier). The original description of the VizieR service was published in A\&AS 143, 23 \citep{Ochsenbein2000}.

This research made use of Montage. It is funded by the National Science Foundation under Grant Number ACI-1440620, and was previously funded by the National Aeronautics and Space Administration's Earth Science Technology Office, Computation Technologies Project, under Cooperative Agreement Number NCC5-626 between NASA and the California Institute of Technology.

The authors made use of the database `astrophysical CATalogues Support system' \citep[CATS;][]{Verkhodanov2005} of the Special Astrophysical Observatory.

Finally, the following open-source software was used for data visualisation and processing: {\sc topcat} \citep{Taylor2005}, SAOImage DS9 \citep{SAO2000, Joye2003}, NumPy \citep{Oliphant2006}, Astropy \citep{Astropy2013}, APLpy \citep{Robitaille2012}, Matplotlib \citep{Hunter2007}, and SciPy \citep{Jones2001}.

\end{acknowledgements}

\begin{appendix}

 
\end{appendix}

\bibliographystyle{pasa-mnras}
\bibliography{SarahWhite_GLEAM_4Jy_DefinitionPaper}

\end{document}